\documentclass[
aps,
nofootinbib,
prd,
superscriptaddress,
tightenlines,
notitlepage,
twocolumn,
showpacs,
floatfix
]{revtex4-1}
\usepackage{amsmath}
\usepackage{latexsym}
\usepackage{amsfonts}
\usepackage{graphicx}
\usepackage{mathrsfs}
\usepackage{epstopdf}
\usepackage{mathtools}
\usepackage{hhline}
\usepackage[utf8]{inputenc}
\usepackage{CJK}
\usepackage{float}
\usepackage{color}
\usepackage{hyperref}
\usepackage{diagbox}
\usepackage{xcolor}
\usepackage{bm}
\usepackage{lipsum}
\usepackage[T1]{fontenc}
\usepackage[utf8]{inputenc}
\usepackage[normalem]{ulem}
\hypersetup{
    colorlinks=true,
    linkcolor=red,
    citecolor=blue,
}

\newcommand\dd{{\rm d}}
\newcommand\bw{\begin{widetext}}
\newcommand\ew{\end{widetext}}

\newcommand{\hhbar}{{ \bar{\xi} }}

\begin{document}
\title{Resumming Kerr Quasinormal-Mode Frequencies: \\ Accuracy and Breakdown Near Extremality}

\author{Jierui Hu}
\email{jieruih2@illinois.edu}
\affiliation{Illinois Center for Advanced Studies of the Universe \& Department of Physics,
 University of Illinois Urbana-Champaign, Urbana, Illinois 61801, USA}

\author{Kent Yagi}
\affiliation{Department of Physics, University of Virginia, Charlottesville, Virginia 22904, USA}
 
\author{Nicol\'as Yunes }
\affiliation{Illinois Center for Advanced Studies of the Universe \& Department of Physics,
 University of Illinois Urbana-Champaign, Urbana, Illinois 61801, USA}
\date{\today}

\begin{abstract}

Kerr black-hole quasinormal modes are usually computed with numerical methods, but analytic approximations remain useful for identifying the physics that controls different parts of the spectrum. In this paper, we ask whether the divergent, high-order Wentzel-Kramers-Brillouin (WKB) expansion about the peak of the Chandrasekhar-Detweiler potential can be made predictive through Pad\'e and Borel-Pad\'e resummation. We develop two complementary implementations: a semi-analytic slow-rotation expansion in the dimensionless spin $a$ (carried out through 21st WKB order), and a fixed-spin Pad\'e-WKB implementation for the resummed frequency equation (carried out through 41st WKB order). In the slow-rotation regime, the 21st-order resummed expansion is significantly more accurate than the fourth-order approximation found previously. For damped modes at larger spins, the fixed-point iteration agrees well with Leaver's method, reaching fractional errors below $10^{-7}$ in the real part of the fundamental $m=0$ mode at $a=0.99$. The same strategy fails for modes that approach the zero-damped branch near extremality. We trace this breakdown to the near-horizon structure of the Chandrasekhar-Detweiler potential. As the extremal limit is approached, nearby poles produce rapid variation on the throat scale, so a local Taylor expansion about the potential peak no longer uniformly captures the relevant region.

\end{abstract}

\maketitle

\section{Introduction}
\label{sec:introduction}

Black-hole quasinormal modes (QNMs) are the damped oscillations of the linearized remnant spacetime that dominate the late-time ringdown after the prompt, nonlinear post-merger response. The ringdown signal in GW150914~\cite{LIGOScientific:2016lio} was consistent with the least-damped Kerr QNM of the remnant black hole, while GW250114~\cite{LIGOScientific:2025wao}, the loudest gravitational-wave signal observed to date, provides a much higher signal-to-noise view of the ringdown regime. Next-generation ground-based detectors~\cite{Branchesi:2023mws,Evans:2021gyd} and future space-based detectors, such as LISA~\cite{LISA:2024hlh}, should measure black-hole QNM frequencies with higher precision and across a broader population of sources. These measurements will provide stringent tests of general relativity and a clean probe of strong-field, dynamical gravity~\cite{Berti:2016lat,Berti:2009kk,Berti:2025hly}.

The calculation of Kerr black-hole QNM frequencies has a long history. On the numerical side, the standard method is Leaver's continued fractions~\cite{Leaver:1985ax}, which determines the QNM frequencies and angular separation constants by solving the radial and angular continued-fraction equations simultaneously. Leaver's method can be further refined by solving the angular sector spectrally~\cite{Cook:2014cta}.
On the analytic side, the most widely used approximation is the Wentzel--Kramers--Brillouin (WKB) method~\cite{Schutz:1985km,Iyer:1986np,Seidel:1989bp}, which expands the effective potential around its peak and solves perturbatively for the QNM frequency using the Taylor coefficients of the expansion. In this context, WKB order refers to the truncation order of the formal expansion obtained after Taylor-expanding the effective potential about its peak. The $N$th WKB order formula keeps $N$ terms in this expansion and depends on derivatives of the potential through order $2N$ at the peak.  

In the black-hole context, the WKB method was first developed for nonrotating black holes~\cite{Schutz:1985km,Iyer:1986np}, and later extended to Kerr black holes by Seidel and Iyer~\cite{Seidel:1989bp} in a slow-rotation expansion, and by Kokkotas~\cite{Kokkotas:1991vz} using a more numerical approach. The early Kerr calculations were carried out to third WKB order, and Kokkotas tabulated only the $(l,m,n)=(2,0,0)$ mode. More recently, Tang et al.~\cite{Tang:2025qaq} revisited Kerr QNMs through fourth WKB order and compared the results with continued-fraction data. For Schwarzschild black holes, the WKB approximation was extended to fifth order by Konoplya~\cite{Konoplya:2003ii}, but the ordinary WKB series is asymptotic, so increasing the order does not by itself guarantee convergence to the numerical result.

\begin{figure*}[tbh]
    \centering
    \includegraphics[scale=0.4]{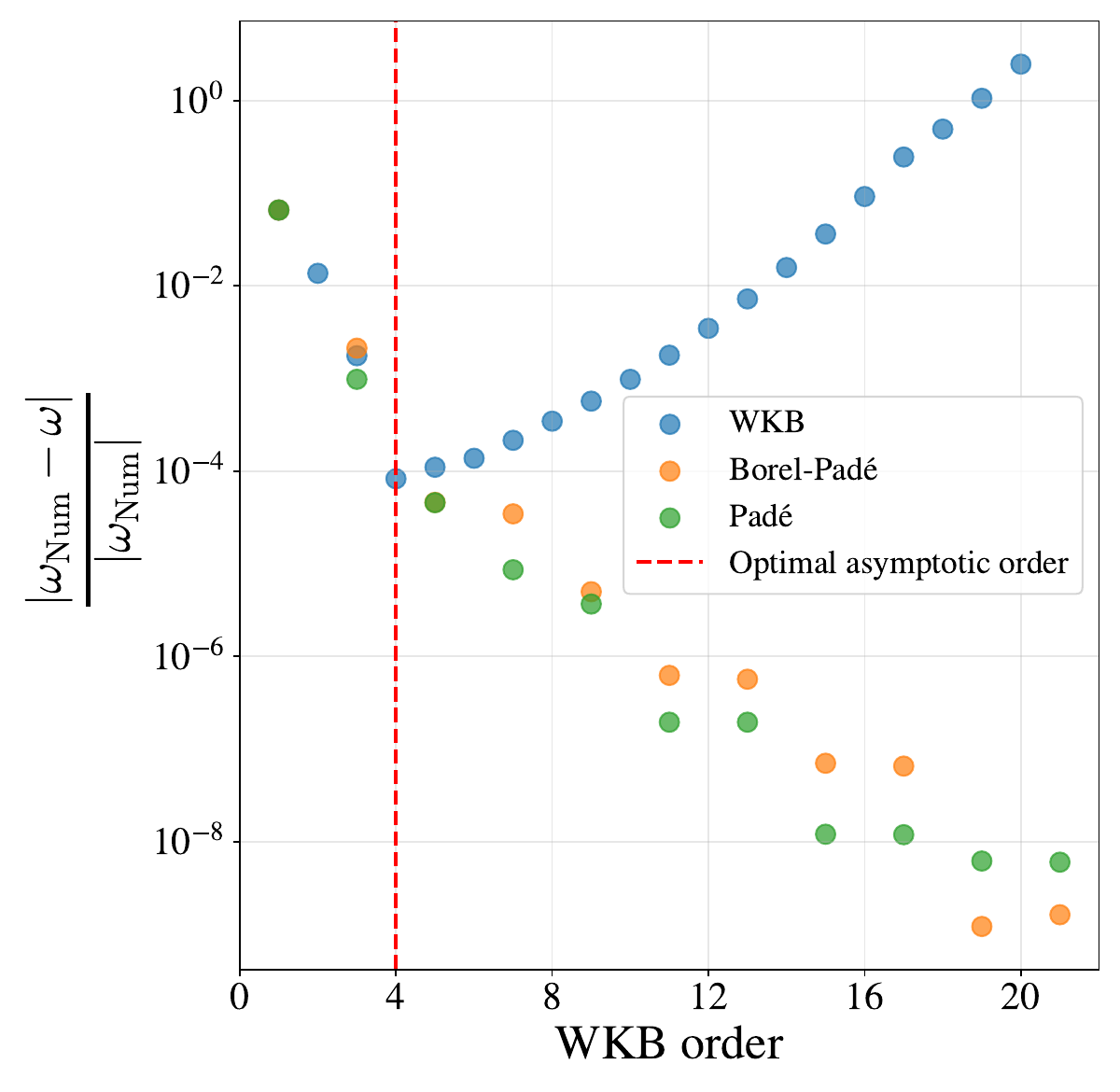}
    \includegraphics[scale=0.53]{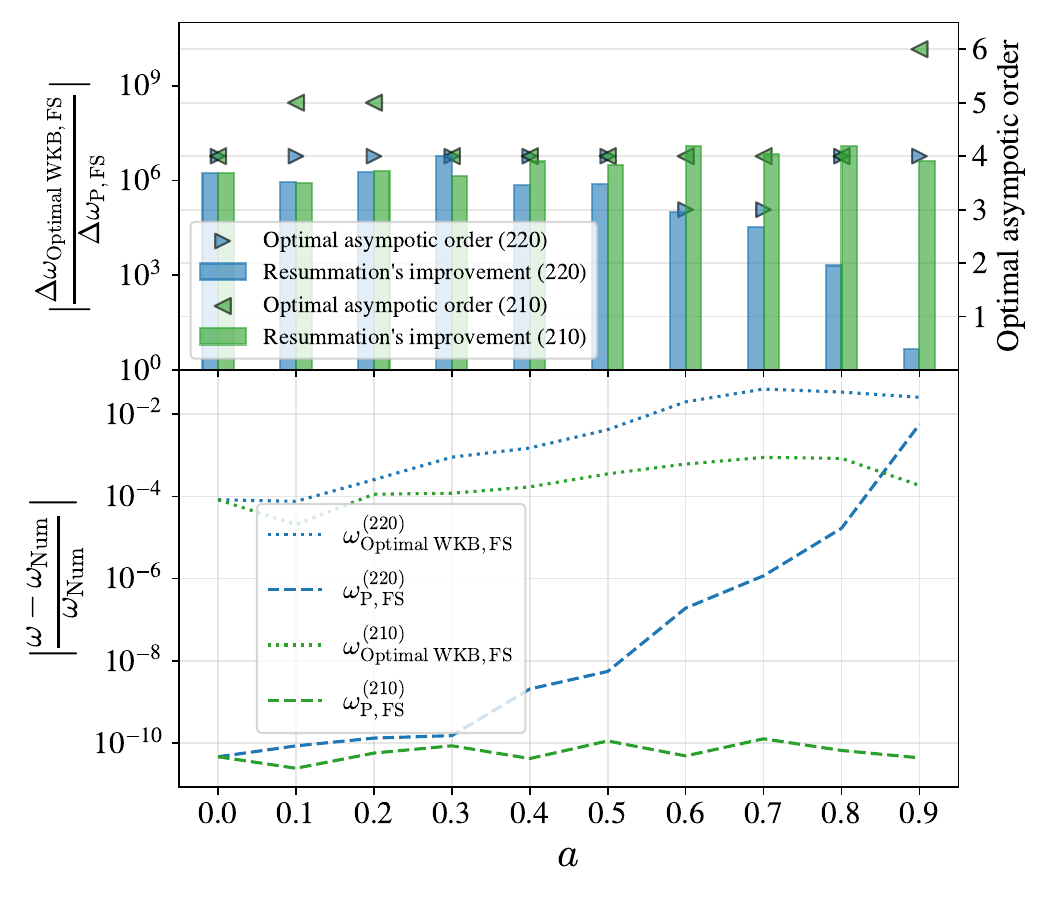}
    \caption{Left: accuracy of different approximation schemes when computing the fundamental $(l,n) = (2,0)$ QNM frequency of a Schwarzschild black hole, relative to numerical results~\cite{Berti:2009kk}. The optimal asymptotic order of the WKB series, $N_{\rm opt}=4$, is indicated by a red dashed line. Pad\'e and Borel-Pad\'e resummation greatly improve the accuracy of the WKB series. Right: Comparison between the WKB approach at the optimal asymptotic order and the 41st-order Pad\'e-resummed WKB approximation (blue/green symbols stand for (l,m,n)=(2,2,0)/(2,1,0) mode). 
    The triangles in the top panel show the optimal asymptotic order as a function of spin. 
    The dotted lines in the bottom panel are the fractional error of the ordinary WKB method at the optimal asymptotic order, while the dashed lines are the fractional error of the 41st-order Pad\'e-resummed WKB. The bars in the top panel show the improvement from the resummation, which is defined as $|\frac{\Delta \omega_{\rm Optimal\,WKB,FS}}{\Delta \omega_{\rm P,FS}}|=|\frac{\omega_{\rm Optimal\,WKB,FS}-\omega_{\rm Num}}{\omega_{\rm P,FS}-\omega_{\rm Num}}|$. Observe that the resummation improves the fractional accuracy of the ordinary WKB approximation at the optimal asymptotic order by more than $10^3$ for $a\leq0.8$ for the $(220)$ model and by about $10^6$ for the $(221)$ mode.
}
    \label{fig:Sch WKB l2}
\end{figure*}

For Schwarzschild black holes, increasing the WKB order initially improves the fundamental-mode frequency, which is why low-order WKB methods are useful. This improvement, however, does \textit{not} imply convergence of the unresummed series. High-order studies showed that the ordinary WKB series is asymptotic~\cite{Konoplya:2019hlu}: after an optimal truncation order, adding more terms eventually worsens the agreement with numerical results~\cite{Matyjasek:2019eeu,Hatsuda:2019eoj}.
We illustrate this behavior in the left panel of Fig.~\ref{fig:Sch WKB l2} for the $(l,n)=(2,0)$ Schwarzschild mode, where the ordinary WKB approximation improves up to $N_{\rm opt}=4$ and then moves away from the numerical value.
Pad\'e and Borel-Pad\'e resummations can instead use the high-order WKB data to recover highly accurate Schwarzschild QNM frequencies~\cite{Matyjasek:2017psv,Matyjasek:2019eeu,Hatsuda:2019eoj}, as we also show in this figure. Motivated by this success, we ask whether analogous resummations can be extended to Kerr black holes, and where such an extension ceases to be accurate.

In this paper, we treat the high-order WKB expansion for Kerr QNM frequencies as an asymptotic series and ask whether its high-order information can be made predictive by resummation. For Schwarzschild black holes, the asymptotic character of the ordinary WKB series is already known. Our Kerr calculations support the same practical lesson: increasing the unresummed WKB order is not, by itself, a controlled route to higher accuracy.

We therefore construct high-order resummations of the Kerr WKB series using the Chandrasekhar-Detweiler potential~\cite{Chandrasekhar:1976zz,Chandrasekhar_1983}. In the slow-rotation regime, we expand the potential, the angular separation constant, the peak location, and the QNM frequency in the spin, compute the WKB series through 21st order, and resum the WKB series with Pad\'e and Borel-Pad\'e approximants. This gives analytic and semi-analytic frequency expansions for the fundamental $l=2$ modes in the Pad\'e and Borel-Pad\'e cases respectively, improving substantially over the ordinary fourth-order WKB approximation in the slow-spin regime.

The end result of the above approaches is still a finite Taylor series in the spin, which is not appropriate to model black holes with moderate rotations. Therefore, we apply a second Pad\'e resummation to the spin expansion itself. This double resummation (Pad\'e or Borel-Pad\'e to the WKB series, followed by Pad\'e to the spin seires) improves the agreement relative to Leaver's method at moderate spins by about an order of magnitude for the $(l,m,n)=(2,0,0)$ mode when $a>0.4$. Nontheless, even this second resummation is not enough to model rapidly rotating black holes in general.

For that reason, we also develop a fixed-spin fixed-point implementation. At fixed $a$, the frequency enters both the Chandrasekhar-Detweiler potential and the angular separation constant, so the resummed WKB condition is no longer an explicit formula for the QNM frequency. Instead, one starts from a trial frequency, computes the corresponding angular separation constant, locates the peak of the Chandrasekhar-Detweiler potential, evaluates the WKB coefficients at that peak, and uses the Pad\'e-resummed WKB condition to update the frequency. Requiring the updated frequency to agree with the trial frequency gives the fixed-point problem, which we solve numerically at each spin and carry out through 41st WKB order. For damped-mode branches, this method can be highly accurate. For example, for the $(l,m,n)=(2,0,0)$ mode, the fractional error in the real part relative to Leaver's method remains below $10^{-7}$ through $a=0.99$.

This accuracy, however, does not extend to modes that approach the zero-damped-mode branch near extremality. For the $(2,2,0)$ mode, the fractional error exceeds $10^{-3}$ for $a>0.9$, signaling the breakdown of the local WKB expansion in this regime.
We trace this breakdown to the near-horizon structure of the Chandrasekhar-Detweiler potential, rather than to the choice of resummation. For branches that approach the zero-damped-mode family, the near-extremal frequency shift scales as $\sqrt{1-a}$, and poles of the Chandrasekhar-Detweiler potential approach the outer horizon on the same scale. The potential, therefore, develops rapid variation in the throat region, with $r-r_+=\mathcal{O}(\sqrt{1-a})$. A Taylor expansion about the potential peak, even when resummed, cannot uniformly approximate both this near-horizon region and the outer region as extremality is approached. This throat-scale structure is the same one that appears in the near-horizon extremal Kerr geometry~\cite{Bardeen:1999px,Amsel:2009ev}; in particular, the right-boundary form of the near-horizon Chandrasekhar-Detweiler potential matches the boundary form of the NHEK radial equation.

Our work should be distinguished from two related recent directions. Tang, Franchini, Völkel, and Berti~\cite{Tang:2025qaq} used low-order WKB methods to estimate QNM frequencies of rotating black holes in GR and beyond GR. That work addresses the practical use of WKB approximations in modified-gravity perturbation equations. Here, instead, we study the high-order asymptotic WKB series for Kerr black holes in GR, ask whether Pad\'e and Borel-Pad\'e resummation can make this series predictive, and identify the near-extremal regime where the resummation breaks down. Our approach is also distinct from exact WKB analyses of black hole QNMs~\cite{Miyachi:2025ptm,Hatsuda:2026ghx}, which resum WKB solutions or quantum periods and impose global quantization conditions through the Stokes structure of the radial equation. We instead resum the local high-order expansion about the peak of the Chandrasekhar-Detweiler potential. In this sense, the present paper isolates what local resummed WKB methods can and cannot do for Kerr QNM frequencies.

The benefit of resumming the WKB series over a wide range of spin is shown in the right panel of Fig.~\ref{fig:Sch WKB l2}.
Although the WKB approach is not ideal for modes near the zero-damped-mode limit, e.g., the (l,m,n)=(2,2,0) mode at $a\geq0.9$, the resummation at the 41st order improves the fractional accuracy of the ordinary WKB approximations at the optimal asymptotic order by more than $10^3$ for $a\leq0.8$ for the $(2,2,0)$ mode, and by abouty $10^6$ for the $(2,2,1)$ mode. 

The remainder of this paper presents the details that led to the conclusions summarized above, and it is organized as follows.
In Sec.~\ref{sec: radial equation}, we revisit the radial master equation and the Chandrasekhar-Detweiler potential.
In Sec.~\ref{sec: basic WKB}, we briefly revisit the resummed WKB techniques.
In Sec.~\ref{Sec: methods}, we present our formulation of the semi-analytic and numerical methods we use to calculate QNM frequencies, based on the resummed WKB approach.
In Sec.~\ref{sec: QNM frequencies}, we report our results for the QNM frequencies of different modes and discuss the validity of the resummed WKB approach for modes with different extremal limits. We draw conclusions in Sec.~\ref{sec: conclusions}. Henceforth, we adopt geometric units, where $G=1=c$, and set the mass of the black hole to $M=1$.

\section{Kerr QNM equations and high-order WKB Resummation} \label{sec:background}

This section fixes the equations and notation used in the rest of the paper. We first review the radial Teukolsky equation for Kerr perturbations and its transformation to the Chandrasekhar-Detweiler potential. We then review the high-order WKB expansion and the Pad\'e and Borel-Pad\'e resummations that will be applied to this potential.

\subsection{Radial Teukolsky equation and Chandrasekhar-Detweiler potential} \label{sec: radial equation}

The Teukolsky equation governs the perturbations of Kerr black holes~\cite{Teukolsky:1973ha}. The radial master equation for a black hole with dimensionless spin $a$ is given by
\begin{align}
    &\Delta^{-s}\frac{d}{dr}\left(\Delta^{s+1}\frac{dR}{dr}\right)+\bigg(\frac{K^2-2 is(r-1)K}{\Delta}
    \nonumber \\ 
    &+4is\omega r-\lambda \bigg)R=0, \label{Radial Teukolsky eq}
\end{align}
where $\omega$ and $s$ are the angular frequency and spin of the perturbed field, and
\begin{equation}
    K=(r^2+a^2)\omega- am, \; \Delta = r^2+a^2-2r,
\end{equation}
\begin{equation}
    \lambda = {}_s{A}_{ml} +a^2\omega^2-2 a m \omega,
  \end{equation}
with ${}_s{A}_{ml}$ the angular separation constant, while $l$ and $m$ are angular quantum numbers. For gravitational perturbation, the spin weight number is $s=-2$.
In the Schwarzschild limit, the angular separation constant has the simple form ${}_s{A}_{ml} = l(l+1)-s(s+1)$. For Kerr black holes, this constant can be solved for numerically from e.g. the continued fraction equation~\cite{Leaver:1985ax} or expanded in a series of small spin:
\begin{equation}
    {}_s{A}_{ml}  = \sum\limits_{p=0}^{N_{A}} f_p (a\omega)^p. \label{eq: Angular separation constant}
\end{equation}
where the first two coefficients are~\cite{Fackerell:1977shn,Seidel:1988ue,Berti:2005gp}
\begin{equation}
\begin{split}
     f_0 &= l(l+1)-s(s+1), \\f_1 &= -\frac{2m s^2}{l(l+1)}\,,
\end{split}
\end{equation}
and the coefficients $f_{p\in (2,6)}$ can be found in~\cite{Fackerell:1977shn,Seidel:1988ue,Berti:2005gp}. 
We will use this series expansion of the angular separation constant when calculating semi-analytic QNM frequencies in a small-spin series (Sec.~\ref{sec: slow-rotation}), and the continued fraction method when working to all spins (Sec.~\ref{sec: iterative WKB}).

The radial Teukolsky equation can be viewed as a Schr\"odinger-like problem with an effective potential, encoded in the second term of Eq.~\eqref{Radial Teukolsky eq}. As we will see in Sec.~\ref{sec: basic WKB}, the WKB approximation requires a well-defined peak of this potential. The difficulty is that the Teukolsky potential remains complex even in the Schwarzschild limit, so its peak is not uniquely defined. We therefore use the Chandrasekhar-Detweiler transformation~\cite{Chandrasekhar:1976zz,Chandrasekhar_1983}, which maps the radial equation to a Schr\"odinger-like equation with a real potential for real $\omega$ and real $r$. For QNMs, where $\omega$ is complex, we use the same expression by analytic continuation.

After this transformation~\cite{Chandrasekhar:1976zz,Chandrasekhar_1983}, the radial equation becomes
\begin{equation}
    \left(\frac{d^2}{d r_\star^2}+\omega^2-V(r,a,\omega) \right)Z= 0, \label{real potential eq}
\end{equation}
             where\footnote{Note that $r_\star$ is not the usual tortoise coordinate $r_{\rm tor}$, which is defined as ${d}/{dr_{\rm tor}}=[{\Delta }/({r^2+a^2})] \, {d}/{dr}$.} ${d}/{dr_\star}=({\Delta }/{ \bar{\rho}^{2}})\, {d}/{dr}$.
The wave function $Z$ can be written as
\begin{equation}
    \frac{\Delta^2}{\bar{\rho}^{8}} (\kappa-4 \omega^2 \beta_2+2 i \omega \kappa_2) Z =( y  - T \Lambda_{-}) |\bar{\rho}^2|^{-3/2} R, \label{eq:Z-R relation}
\end{equation}
where 
\begin{align}
 \bar{\rho}^{2}&=r^2+a^2-\frac{a \, m}{\omega}\,,
 \qquad
 \Lambda_{\pm} = \frac{d}{dr_{\star}} \mp i \omega,
\\\
\kappa &=\lambda(\lambda+2),
\quad
\beta_2 = \pm 3\left(a^2-\frac{a \,m}{\omega}\right),\\
\kappa_2 &= \pm \Big\{36+2 \beta_2 \kappa 
 -2\lambda\left[\left(a^2-\frac{am}{\omega}\right)\left(5\lambda+6\right)-12 a^2\right]\Big\}^{1/2},
\end{align}
and
\begin{equation}
\begin{split}
F&=\frac{1}{\Delta}[\lambda \bar{\rho}^4+3 \bar{\rho}^2 (r^2-a^2)-3 r^2 \Delta],\\
y&=\frac{\Delta^2}{\bar{\rho}^8}(F+\beta_2),\\
T&=\frac{F'-\kappa_2}{F-\beta_2}- 2 i \omega, \;\; F'=F_{,r_\star},
\end{split}
\end{equation}
with a comma sub-index, such as ``$F_{,r_\star}$,'' representing a partial derivative, in this example $F_{,r_\star} = \partial F/\partial_{r_\star}$.

The Chandrasekhar-Detweiler potential $V$ is a function of $r$, $a$, and $\omega$, given by
\begin{equation}
V=-\frac{\Delta^2}{\bar{\rho}^8} \beta_2+\frac{\kappa}{F+\beta_2}-\frac{\left(F^{\prime}-\kappa_2\right)\left(\kappa_2 F-\beta_2 F^{\prime}\right)}{\left(F-\beta_2\right)\left(F^2-\beta_2^2\right)} .
\end{equation}
There are four different choices of $V$ depending on the signature choices of $\beta_2$ and $\kappa_2$. 
Nevertheless, the QNM frequencies derived from Eq.~(\ref{real potential eq}) with these four different choices are the same, because they all satisfy Eq.~(\ref{Radial Teukolsky eq}). In this work, we adopt plus signs for both $\beta_2$ and $\kappa_2$.
In the Schwarzschild limit, when $a=0$, the Chandrasekhar-Detweiler potential reduces to the Zerilli~\cite{Zerilli:1970se} potential when choosing a positive sign for $\kappa_2$, while it reduces to the Regge–Wheeler~\cite{Regge:1957td} potential when choosing a negative sign (notice that $\beta_2 = 0$ in the $a = 0$ limit).

\subsection{High-order WKB expansion and Pad\'e/Borel-Pad\'e resummation} \label{sec: basic WKB}

The Schr\"odinger-like form of Eq.~\eqref{real potential eq} allows us to apply the WKB approximation to the radial problem. In this approach, one introduces a formal small parameter to organize a local asymptotic expansion of the wave equation near the peak of the effective potential. The approximation is useful when the QNM frequency is controlled primarily by this barrier region, so that the potential is smooth and single-peaked and the relevant turning points lie close enough to the peak for its Taylor expansion to capture the connection problem. This expectation is best justified for low overtones and larger angular quantum numbers, and it can fail when additional structure away from the peak becomes important.

To set up this expansion, let us first assume that the potential $V$ depends only on $r$. The Kerr calculation requires the more general case in which $V$ also depends on $a$ and $\omega$, and we will return to this point in Sec.~\ref{sec: slow-rotation}.
We then begin by expanding the potential around its peak $r_\star=r_{\star0}$ as follows:
\begin{equation}
    V(r_\star)= V_0- \sum\limits^{\infty}\limits_{k=2} V_k (r_\star-r_{\star 0 })^k\,,
\end{equation}
where 
\begin{align}
 V_k = \left.- \frac{1}{k!} \frac{d^k V(r_\star)}{d r_{\star} ^k}\right|_{r_\star=r_{\star0}}.   
\end{align}
Following~\cite{Hatsuda:2019eoj}, we redefine
$r_\star-r_{\star0}=\sqrt{i \, \xi}\,\tilde{x}$, and further define
$\hhbar=i \, \xi$, so that $r_\star-r_{\star0}=\sqrt{\hhbar}\,\tilde{x}$.
The quantity $\xi$ is only a bookkeeping parameter, not the reduced Planck constant, so at the end of the calculation we will set $\xi=1$, or equivalently $\hhbar=i$.
With these definitions, we can later carry out expansions in $\xi \ll 1$, which are equivalent to expansions in $\hhbar \ll 1$. Essentially, expanding the potential in $\xi \ll 1$ corresponds to zooming in to its peak along the real line, while expanding in $\hhbar \ll 1$ does the same but along a path in the complex plane. 

We are now ready to recast Eq.~\eqref{real potential eq} in a WKB-amenable form. Multiplying Eq.~(\ref{real potential eq}) by $-1/2$ and substituting $r_\star$ with $\tilde{x}$, we obtain the perturbed harmonic oscillator equation
\begin{align}
    \left(-\frac{1}{2}\frac{\dd^2}{\dd \tilde{x}^2}+\frac{V_2 }{2}\tilde x^2+\frac{1}{2}\sum\limits^{\infty}\limits_{k=3}\hhbar^{k/2-1} \right.& V_k \tilde{x}^k \Biggr)  \psi(\tilde{x})
    \nonumber \\
    &=\left(\frac{V_0-\omega^2}{2\hhbar} \right)\psi(\tilde{x}),
\label{final anharmonic eq}
\end{align}
where the perturbation is anharmonic, and the first derivative of the potential, $V_1$, is assumed to vanish by definition. 
The WKB approximation treats Eq.~(\ref{final anharmonic eq}) as a local asymptotic problem near the peak of the effective potential. At leading order, one keeps only the quadratic term in the Taylor expansion of the potential, which gives the parabolic-barrier approximation. Higher WKB orders incorporate the higher Taylor coefficients $V_3,V_4,\ldots$ as perturbations of this local oscillator problem.

The quantity $(V_0-\omega^2)/(2\hhbar)$ on the right-hand side of Eq.~(\ref{final anharmonic eq}) plays the role of the eigenvalue of the local problem. This eigenvalue can be computed recursively by expanding the wave function and the eigenvalue in powers of $\hhbar$, substituting these expansions into Eq.~(\ref{final anharmonic eq}), and matching equal powers of $\hhbar$. Such an approach is the Bender-Wu recursion~\cite{Bender:1969si}, for which we use the Mathematica implementation of Ref.~\cite{Sulejmanpasic:2016fwr}. The result takes the form
\begin{equation}
    \frac{V_0-\omega^2}{2\hhbar}
    =
    \sum_{k=0}^{N-1}\epsilon_k\hhbar^k
    +\mathcal{O}(\hhbar^N),
\end{equation}
where the coefficients $\epsilon_k$ are fixed by the differential equation and depend on the Taylor coefficients $V_j$ of the potential at the peak. Equivalently, the $N$th-order WKB approximation to the QNM frequency is
\begin{equation}
    \omega^2
    =
    V_0-\left(2\hhbar\sum_{k=0}^{N-1}\epsilon_k\hhbar^k\right)\bigg|_{\hhbar=i}.
\label{eq: WKB omega}
\end{equation}
The $N$th WKB order keeps the coefficients $\epsilon_0,\ldots,\epsilon_{N-1}$ and uses derivatives of the potential through ${\cal{O}}(\hhbar^{2N})$ at the peak.
Writing $\alpha=n+1/2$, where $n$ is the overtone number, the first few coefficients are~\cite{Hatsuda:2019eoj}
 \begin{align}
    \epsilon_0 = & \sqrt{V_2} \alpha,\label{eq: WKB0}\\
    \epsilon_1 =& -\frac{V_3^2}{64 V_2^2}(7+60\alpha^2)+\frac{3V_4}{16 V_2}(1+4\alpha^2), \label{eq: WKB1} \\
    \epsilon_2 =& \alpha\biggl[-\frac{15V_3^4}{1024V_2^{9/2}}(77+188\alpha^2)\nonumber \\
&+\frac{9V_3^2 V_4}{128V_2^{7/2}}(51+100\alpha^2)-\frac{V^2_4}{64V_2^{5/2}}(67+68\alpha^2) \nonumber \\
&-\frac{5V_3 V_5}{32V_2^{5/2}} (19+28\alpha^2)+\frac{5V_6}{16V_2^{3/2}}(5+4\alpha^2)\biggr].\label{eq: WKB2}
\end{align}
These coefficients agree with the third-order WKB coefficients of Ref.~\cite{Iyer:1986np}. When $N=1$, Eq.~(\ref{eq: WKB omega}) reduces to the standard first-order WKB result,
\begin{equation} 
    \omega^2 = V_0  - 2 i\sqrt{V_2} \left(n+\frac{1}{2}\right).
\label{eq:first-order WKB}
\end{equation}

For Schwarzschild black holes, high-order studies show that the unresummed WKB series is asymptotic rather than convergent~\cite{Konoplya:2019hlu,Hatsuda:2019eoj}. As illustrated in Fig.~\ref{fig:Sch WKB l2} for the fundamental $(l,n)=(2,0)$ mode, the ordinary WKB approximation improves up to the optimal order $N_{\rm opt}=4$, reaching a fractional error of order $10^{-4}$, and then worsens at higher order. This accuracy may well be sufficient for current ringdown analyses of the dominant mode, but observational adequacy at a given signal-to-noise ratio is not the same as convergence of the formal WKB series. Moreover, the good behavior of the Schwarzschild fundamental mode should not be taken as representative of all cases: for Kerr black holes, recent low-order WKB comparisons find mode- and spin-dependent accuracy, with better performance for larger $\ell$ and lower $n$, and non-monotonic improvement with WKB order for some low-$\ell$ modes~\cite{Tang:2025qaq}.

In spite of its asymptotic nature, the accuracy of the WKB approximation can often be improved through resummation methods, such as Pad\'e resummation~\cite{Matyjasek:2017psv,Matyjasek:2019eeu} or Borel-Pad\'e resummation~\cite{Hatsuda:2019eoj}. These resummations use the information contained in the high-order WKB coefficients without treating the truncated WKB series as a convergent Taylor series. As shown in Fig.~\ref{fig:Sch WKB l2}, this strategy can yield highly accurate QNM frequencies for Schwarzschild black holes.

Note that the convergence of the resummation is not necessarily monotonic, as shown in Ref.~\cite{Konoplya:2026rjh}. We also observe this non-monotonic convergence in the 11th- and 13th-order Pad\'e resummed results in the left panel of Fig.~\ref{fig:Sch WKB l2}. Nevertheless, the Ces\'aro means of the resummed series can stabilize the convergence~\cite{Konoplya:2026rjh}.

We now define the resummations used in this work. Let $N_{\rm max}=N-1$ be the highest power of $\hhbar$ kept in the WKB series, and let
\begin{equation}
    N_{\rm P}\equiv \frac{N_{\rm max}}{2},
\end{equation}
where $N_{\rm P}$ is the degree of the numerator and denominator of the diagonal Pad\'e approximant. We only consider even values of $N_{\rm max}$ in this paper. The Pad\'e-resummed WKB approximant to the \textit{square} of the QNM frequencies is then
\begin{equation}
    \omega_{\rm P}^2
    =
    V_0-
    \left[
    2\hhbar\,
    P^{N_{\rm P}}_{N_{\rm P}}
    \left(\sum_{k=0}^{N_{\rm max}}\epsilon_k\hhbar^k\right)
    \right]_{\hhbar=i},
\label{eq: Pade WKB frequencies}
\end{equation}
where
\begin{align}
P^{N_{\rm P}}_{N_{\rm P}}
\left(\sum_{k=0}^{N_{\rm max}}\epsilon_k\hhbar^k\right)
=
\frac{
p_0+p_1\hhbar+\cdots+p_{N_{\rm P}}\hhbar^{N_{\rm P}}
}{
1+q_1\hhbar+\cdots+q_{N_{\rm P}}\hhbar^{N_{\rm P}}
}.
\end{align}
The coefficients $p_i$ and $q_i$ are fixed by requiring the Taylor expansion of this rational function to agree with the WKB series through order $N_{\rm max}$:
\begin{align}
\frac{
p_0+p_1\hhbar+\cdots+p_{N_{\rm P}}\hhbar^{N_{\rm P}}
}{
1+q_1\hhbar+\cdots+q_{N_{\rm P}}\hhbar^{N_{\rm P}}
}
-
\sum_{k=0}^{N_{\rm max}}\epsilon_k\hhbar^k
=
\mathcal{O}(\hhbar^{N_{\rm max}+1}) .
\end{align}
The Borel-Pad\'e approximant is constructed similarly, but after first taking the Borel transform of the WKB series. In our notation, this gives
\begin{equation}
\begin{split}
    \omega_{\rm BP}^2
    =
    V_0
    -
    \left\{
    2\hhbar
    \int^\infty_0 e^{-\zeta}
    P^{N_{\rm P}}_{N_{\rm P}}
    \left[
    \sum_{k=0}^{N_{\rm max}}
    \frac{\epsilon_k}{k!}
    (\zeta \hhbar)^k
    \right]
    d\zeta
    \right\}_{\hhbar=i},
\label{eq: Borel-Pade WKB frequencies}
\end{split}
\end{equation}
where $\zeta$ is the Borel integration variable.

The reason to resum the WKB series is that the high-order coefficients still contain useful information even when the ordinary partial sums stop converging. For an asymptotic series, the first few terms may approach the desired value, but the coefficients eventually grow fast enough that adding further terms worsens the approximation. As we can see above, Pad\'e resummation replaces the truncated polynomial in $\hhbar$ by a rational function whose Taylor expansion agrees with the WKB series through the computed order. This rational approximation can capture part of the analytic structure of the underlying function, such as nearby poles or branch-cut behavior, that is invisible in a finite polynomial truncation. 

Similarly, Borel-Pad\'e resummation addresses the same problem in a slightly different way. If the large-order WKB coefficients grow factorially, the Borel transform, obtained by dividing the $k$th coefficient by $k!$, can have a finite radius of convergence even when the original series does not. One then approximates the Borel transform by a Pad\'e approximant and reconstructs the resummed answer through a Laplace-type integral, as shown above. This procedure is known to work well for Schwarzschild QNMs~\cite{Hatsuda:2019eoj}, and it motivates applying the same resummation strategy to the Kerr WKB series, as we do in this paper.  Neither Pad\'e nor Borel-Pad\'e resummation guarantees improvement for every mode or spin; their usefulness must be checked against numerical QNM frequencies, as we do below.

As discussed in Sec.~\ref{sec: slow-rotation}, both the Pad\'e resummed and Borel-Pad\'e resummed approximants yield
similar results for QNM frequencies under a slow-rotation expansion. However, the Borel-Pad\'e approximant, which involves a numerical integration for large values of $N_{\rm max}$, introduces additional numerical uncertainty and is more computationally intensive. 
In contrast, the Pad\'e resummed approximant is purely algebraic, once the WKB coefficients are known. Therefore, in the fixed-spin Pad\'e-WKB calculation of Sec.~\ref{sec: iterative WKB}, we use only the Pad\'e-resummed WKB approximant.

\section{Resummed WKB constructions for Kerr QNM frequencies} \label{Sec: methods}

This section develops two resummed WKB implementations. The first is a semi-analytic slow-rotation construction, in which the potential, angular separation constant, and QNM frequency are expanded in powers of the spin. The second is a fixed-spin numerical construction, in which the Pad\'e-resummed WKB equation is solved iteratively for the frequency.

\subsection{Black holes with small spins} \label{sec: slow-rotation}

In this subsection, we solve for QNM frequency solutions under the slow-rotation expansion up to the $N_a$-th order, which is
\begin{equation}
    \omega(a)=\omega_0 +\omega_1 a+ \omega_2 a^2 +...+\omega_{N_a}  a^{N_a} + \mathcal{O}(a^{N_a+1}), \label{eq: omega series of a}
\end{equation}
where each coefficient, $\omega_i$, can be obtained analytically from $r_{0, \rm{Sch}}$, which denotes the peak location of the potential for the Schwarzschild black hole. 
In our signature convention (see Sec.~\ref{sec: radial equation}), the potential reduces to the Zerilli potential in the Schwarzschild limit, so $r_{0, \rm{Sch}} = 3.0987906442...$.

The steps to obtain $\omega_i$ are as follows.

\begin{enumerate}
    \item \textbf{Find the peak of the potential.}
   
To apply the WKB approximation under the slow-rotation expansion, we begin by finding the peak location in a slow-rotation expansion
\begin{align}
    r_{0}(\omega,a) &= r_{0, \rm{Sch}}+ r_{0,1}(\omega) a^1+ r_{0,2} (\omega) a^2+ \ldots \nonumber 
    \\&+ r_{0,N_a} (\omega) a^{N_a}  + \mathcal{O}(a^{N_a+1}), \label{eq: r expanded in a}
\end{align}
where the coefficients $r_{0,i}(\omega), \;(i\neq0)$ can be obtained analytically by solving the equation 
\begin{equation}
    V'(r,a,\omega)=\frac{d V(r,a,\omega)}{d r} = 0,
\end{equation}
order by order in $a$.
The angular separation constant $\lambda$ in $V(r,a,\omega)$ is also expanded up to $N_a$-th order in $a$.

\vspace{3mm}

We solve for Eq.~(\ref{eq: r expanded in a}) in the following manner. At the first order in $a$, $V'(r,a,\omega)=0$ is a degree-one polynomial  for $r_{0,1}$, allowing us to express $r_{0,1}$ as a function of $\omega$. Similarly, at the $N_a$-th order in $a$, with the coefficients $r_{0, i},\; (i<N_a)$ and $r_{0, \rm{Sch}}$ already determined, $V'(r,a,\omega)=0$ reduce to a degree-one polynomial for $r_{0,n}$. This yields $r_{0,n}$ as a function of $\omega$.

\item \textbf{Evaluate the derivatives of the potential $V$ at the peak.}

Given the series expansion of $r_{ 0 }$ in terms of $a$, we can expand $V_k(\omega,a) = -{d^k V(r,a,\omega)}/({k! d r_{\star} ^k})|_{r=r_{0}}$ in a series of $a$, namely
\begin{align}
    V_k(\omega,a) &=  V_{k,0}+V_{k,1}(\omega) a+V_{k,2}(\omega) a^2+\ldots
    \nonumber \\
    &+V_{k,N_a}(\omega) a^{N_a}+ \mathcal{O}(a^{N_a+1}), \label{eq: Vk expanded}
\end{align}
where 
$V_{k,q}(\omega)$ denotes the coefficient of $a^q$ in the slow-rotation expansion of the $k$th derivative of the potential at the peak:
\begin{equation}
    V_{k,q}(\omega)
    =
    [a^q]\left\{
    -\frac{1}{k!}
    \left[
    \left(
    \frac{\Delta}{\bar{\rho}^2}\frac{d}{dr}
    \right)^k
    V(r,a,\omega)
    \right]_{r=r_0(\omega,a)}
    \right\}.
\end{equation}
Here $[a^q]\{\cdots\}$ means ``take the coefficient of $a^q$'' after substituting the slow-rotation expansions for the peak location $r_0(\omega,a)$ and the angular separation constant. This definition includes the explicit spin dependence of $V$, the spin dependence of the tortoise derivative $d/dr_\star=(\Delta/\bar{\rho}^2)d/dr$, and the implicit spin dependence through $r_0(\omega,a)$.

\item \textbf{Pad\'e and Borel-Pad\'e resum the WKB series.}

Using the recursion relations from \cite{Bender:1969si,Sulejmanpasic:2016fwr} and Eq.~(\ref{eq: Vk expanded}), we can expand the WKB coefficients $\epsilon_k(\omega,a)$ in a small-spin series, namely
\begin{align}
    \epsilon_k(\omega,a) &= \epsilon_{k,0}+\epsilon_{k,1}(\omega) a+\epsilon_{k,2}(\omega) a^2+\ldots \nonumber\\&+\epsilon_{k,N_a}(\omega) a^{N_a}+ \mathcal{O}(a^{N_a+1}).
\end{align}
Since $a$ appears in the Chandrasekhar-Detweiler potential with a factor of either $\omega$ or $1/\omega$, the dependence of $\epsilon_{k,n}(\omega)$ can be expressed as a polynomial in $\omega$ and $1/\omega$ of degree at most $N_a$:
\begin{equation}
    \epsilon_{k,q}(\omega) = \sum\limits_{i=-N_a}^{i=N_a}  \epsilon_{k,q,i} \; \omega^{i},
\end{equation}
where $\epsilon_{k,q,i}$ are constants that are independent of $a$ and $\omega$. The $k$ sub-index stands for the derivative order of the potential, while the $q$ sub-index counts powers of $a$ and the $i$ sub-index powers of $\omega$.

\vspace{3mm}

Thus, the resummations in Eqs.~(\ref{eq: Pade WKB frequencies}) and~(\ref{eq: Borel-Pade WKB frequencies}) can be modified into
\begin{align}
    \omega_{\rm P,Taylor}(a)^2 &= V_0- \sum\limits_{j=0}^{N_a}\sum\limits_{i=-j}^{i=j} a^j \; \omega_{\rm P,Taylor}(a)^{i} 
    \nonumber \\
    &\times \left[2 \, \hhbar \,  P^{N_{\rm P}}_{N_{\rm P}}\left(\sum_{k=0}^{N_{\rm{max}}} \epsilon_{k,j,i} \hhbar^k\right)\right]\Bigg|_{\hhbar=i}, \label{eq:Pade slow ro}
\end{align}
and
\begin{align}
    &\omega_{\rm BP,Taylor}(a)^2 = V_0
    - \sum\limits_{j=0}^{N_a}\sum\limits_{i=-j}^{j} a^j \omega_{\rm BP,Taylor}(a)^{i}
    \nonumber \\& 
    \times \left\{2 \hhbar  \int^\infty_0 e^{-\zeta} P^{N_{\rm P}}_{N_{\rm P}}\left[\sum_{k=0}^{N_{\rm{max}}} \frac{\epsilon_{k,j,i}}{k!} (\zeta \hhbar)^k\right] d\zeta \right\}\Bigg|_{\hhbar=i}. \label{eq:Borel-Pade slow ro}
\end{align}

In Eqs.~(\ref{eq:Pade slow ro}) and~(\ref{eq:Borel-Pade slow ro}), the resummation is applied coefficient by coefficient in the joint expansion in $a$ and $\omega$. This defines our semi-analytic slow-rotation approximation.

\item \textbf{Calculate the QNM frequencies in the slow-rotation limit.}

The equations above are implicit for the QNM frequency, because the right-hand sides still contain powers of $\omega(a)$. To solve them in the slow-rotation expansion, we substitute the ansatz in Eq.~(\ref{eq: omega series of a}) into either Eq.~(\ref{eq:Pade slow ro}) or Eq.~(\ref{eq:Borel-Pade slow ro}), move all terms to one side, and expand the result in powers of $a$. Requiring the coefficient of each power of $a$ to vanish then determines the coefficients $(\omega_0,\omega_1,\ldots,\omega_{N_a})$ recursively. This gives the QNM frequency as a Taylor series in the spin:
\begin{equation}
    \omega_{\rm X,Taylor}(a)=\sum_{j=0}^{N_a}\omega_{j,\rm X} a^j,
    \label{eq:first-resummation}
\end{equation}
but with the $\hhbar$ series Pad\'e-resummed (X=P) or Borel-Pad\'e-resummed (X=BP).

 \end{enumerate}

In Appendix~\ref{app:first-order-pade-example} we provide a concrete example of this approach, so that our methodology is clear. 
%

\subsection{Black holes with moderate and rapid spins}
\label{sec: iterative WKB}
\begin{figure}[tbh]
    \centering
    \includegraphics[scale=0.4]{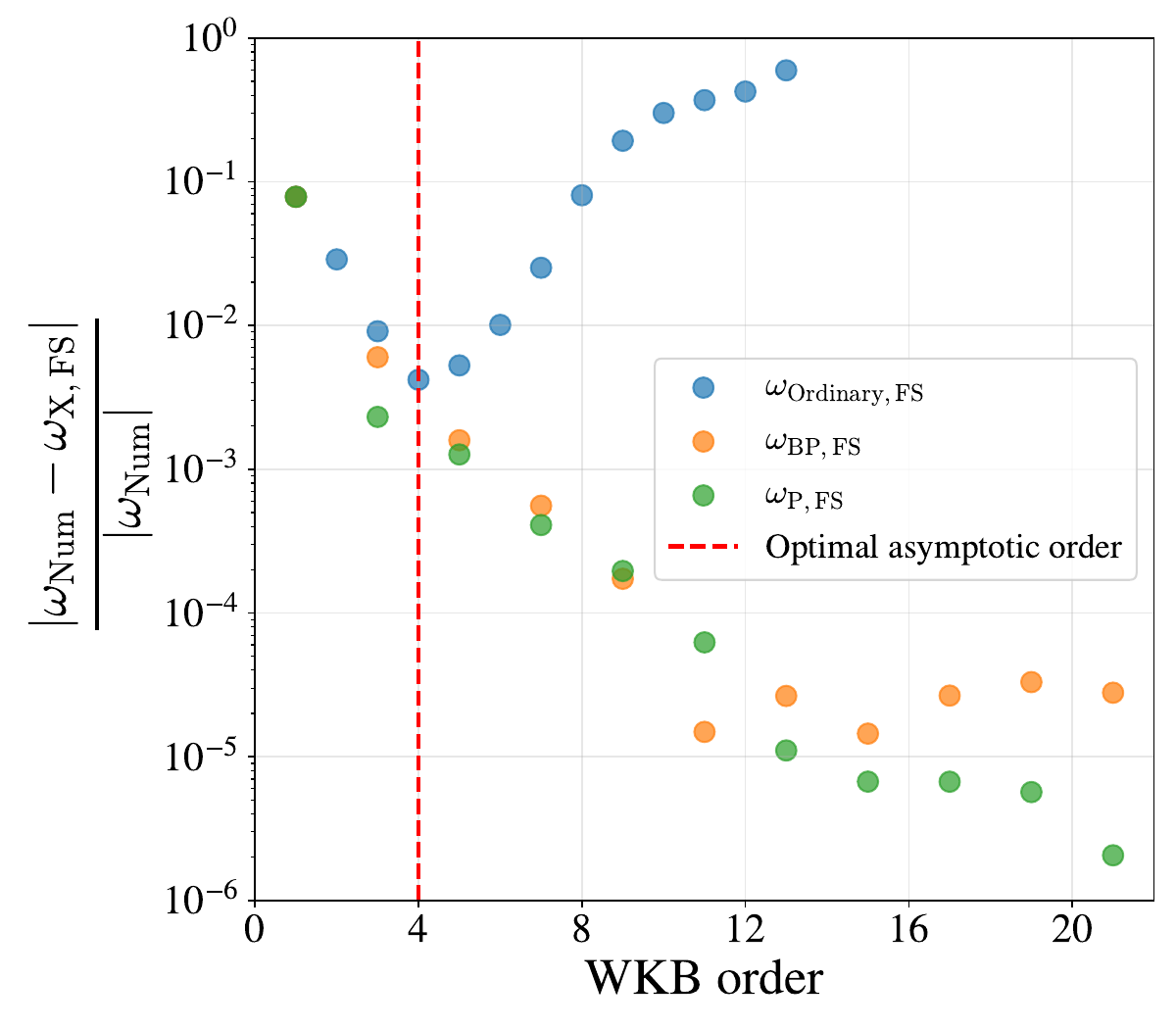}
    \caption{Accuracy of different fixed-spin approximation schemes when computing the fundamental $(l,m,n) = (2,2,0)$ QNM frequency of a Kerr black hole with $a=0.5$, relative to numerical results~\cite{Berti:2009kk}. The optimal asymptotic order of the WKB series, $N_{\rm opt}=4$, is indicated by a red dashed line. Both resummations still greatly improve the ordinary WKB series. However, Pad\'e resummation outperforms Borel-Pad\'e resummation when the WKB order exceeds 12, which is attributed to additional numerical error introduced by the Borel integration.}
    \label{fig: Kerr WKB 220 a05}
\end{figure}

Let us now extend the resummed WKB approach beyond the slow-rotation expansion in two ways. The results of the previous subsection are still a finite Taylor series in the spin $a$, as shown in Eq.~\eqref{eq:first-resummation}.  To improve the behavior of this slow-rotation expansion at moderate spins, a first option is to apply a second Pad\'e resummation, now with respect to $a$. We thus define
\begin{equation}
    \omega_{\rm X,P}(a)
    =
    P^{N_a/2}_{N_a/2}
    \left[
    \sum_{j=0}^{N_a}\omega_{j,\rm X} a^j
    \right],
    \label{eq: additional pade}
\end{equation}
where $X={\rm P}$ or $X={\rm BP}$, depending on whether the $\hhbar$ series has been Pad\'e resummed or Borel-Pad\'e resummed. Thus, $\omega_{\rm P,P}$ is the Pad\'e-resummed WKB series with a second Pad\'e resummation of the spin series, while $\omega_{\rm BP,P}$ is the Borel-Pad\'e-resummed WKB series with a second Pad\'e resummation of the spin series. This second Pad\'e approximant resums the slow-rotation series, not the WKB series. We provide a concrete example of this approach in Appendix~\ref{app:first-order-pade-example}.

We will see, however, that this second Pad\'e resummation is still not enough to model the QNM frequencies of rapidly rotating black holes, so we develop a second approach: a fixed-point strategy.  
By fixing the value of $a$ to some number, the resummed WKB approach in Eq.~\eqref{eq: Pade WKB frequencies} provides an approximate formula\footnote{For simplicity, we only employ Pad\'e resummation in this subsection, as mentioned at the end of Sec.~\ref{sec: basic WKB}.} for $\omega$
\begin{align}
\label{eq: fixed point}
    \omega &= \Omega(\omega) 
    \nonumber \\
    &:= \sqrt{V_0(\omega)- \left[2\hhbar  P^{N_{\rm{max}}/2}_{N_{\rm{max}}/2}\left(\sum_{k=0}^{N_{\rm{max}}} \epsilon_k (\omega) \hhbar^k\right)\right]|_{\hhbar=i}},
\end{align}
which is a numerical function of $\omega$.
For the fixed-point/fixed-spin iterative method, we use only the Pad\'e resummed WKB expression. Thus, the high-spin iterative results below should be interpreted as Pad\'e resummed WKB results, whereas the slow-rotation results allow a direct comparison between Pad\'e and Borel-Pad\'e resummation.
The coefficients $V_k(\omega)$ and $\epsilon_k(\omega)$ are obtained in the following ways.

Given a specific value of $a$ and an input value of $\omega$\footnote{This input value may not be the "true" value of $\omega$. }, one can find the peak location of the potential $r_0(\omega)$ by solving ${d V(r,a,\omega)}/{d r} = 0$ numerically, where the angular separation constant $\lambda$ is obtained via continued fractions~\cite{Leaver:1985ax,Stein:2019mop}. Once $r_0(\omega)$ is numerically determined, the potential derivatives $V_k(\omega)$ can be calculated from their definition $V_k(\omega)= -{d^k V(r,a,\omega)}/({k! d r_{\star} ^k})|_{r=r_{0}(\omega)}$, while $\epsilon_k(\omega)$ can be evaluated using the recursion relations presented in~\cite{Bender:1969si,Sulejmanpasic:2016fwr}.

The problem of determining the QNM frequency \(\omega\) can be framed as solving a fixed point problem, defined by Eq.~(\ref{eq: fixed point}). In numerical analysis, various methods exist for finding fixed points. One straightforward approach is the fixed-point iteration:
\begin{equation}
    \omega_{j} = \Omega(\omega_{j-1}) = \Omega(\Omega(\omega_{j-2}))=...  \; ,
\end{equation}
However, the fixed-point iteration can diverge when the derivative $ |\Omega'(\omega)|>1$ at the fixed points. To ensure our iterations converge for a large range of $a$ values, we here use the secant method:
\begin{equation}
    \omega_{j} = \omega_{j-1}-\frac{\left(\Omega( \omega_{j-1})-\omega_{j-1}\right)(\omega_{j-1}-\omega_{j-2})}{\Omega( \omega_{j-1})-\omega_{j-1}-\Omega( \omega_{j-2})+\omega_{j-2}}. \label{eq: secant iteration}
\end{equation}
To find the QNM frequency $\omega$ for a specific value of $a$, we start with an initial guess $\omega_0$, chosen either from the Schwarzschild frequency or from the frequency at a nearby value of $a$. We generate the second initial value $\omega_1$ by imposing a $\sim1\%$ deviation from $\omega_0$, and then iterate Eq.~(\ref{eq: secant iteration}) until the result stabilizes. We denote the frequencies obtained in this way by $\omega_{\rm P,FS}(a)$, where ${\rm P}$ indicates Pad\'e resummation of the WKB series and ${\rm FS}$ indicates that the frequency is obtained at fixed spin.

The advantage of the resummed WKB series over the ordinary WKB series under the iterative fixed-spin approach is shown in Fig.~\ref{fig: Kerr WKB 220 a05}, where different approximation schemes are compared to the numerical results using Leaver's method.
We denote the frequencies obtained from the ordinary un-resummed WKB approximation under the fixed-spin method by $\omega_{\rm Ordinary,FS}$, and the Borel-Pad\'e resummed as $\omega_{\rm BP,FS}$. 
The optimal asymptotic order of the ordinary WKB approximation is still 4. 
The improvements by resumming the WKB series are more significant at the 3rd and 5th WKB order, compared to the non-rotating results in Fig.~\ref{fig:Sch WKB l2}.
However, the Borel-Pad\'e resummed WKB approach has larger errors than the Pad\'e resummed WKB approach above the 12th WKB order, which is due to the numerical error introduced by the Borel integration, and this error is further amplified by the iterations around the fixed point. 
Therefore, we will use the Pad\'e resummation only for the fixed-spin approach in Sec.~\ref{sec: iterative WKB frequencies}.

The resummed WKB iteration described above may not always converge to a stable value for all modes $(l,m,n)$ and spins $a$. This method can fail when the potential cannot be accurately represented by a Taylor expansion, which is the foundation of the high-order WKB analysis. For example, in the near-extremal regime, the Taylor-expanded potential for the $(l,m,n)=(2,2,0)$ mode does not approximate the true potential very well (see Sec.~\ref{sec: iterative WKB frequencies}). This mode has a special extremal limit known as zero-damped modes, which is associated with the near-horizon geometry of near-extremal black holes. The near-horizon structure of the Chandrasekhar-Detweiler potential is examined in Sec.~\ref{sec: iterative WKB frequencies}. Further details on zero-damped modes can be found in~\cite{Yang:2012pj,Yang:2013uba}.
\begin{figure*}[!ht]
    \centering
    \includegraphics[clip=true,width=\textwidth]{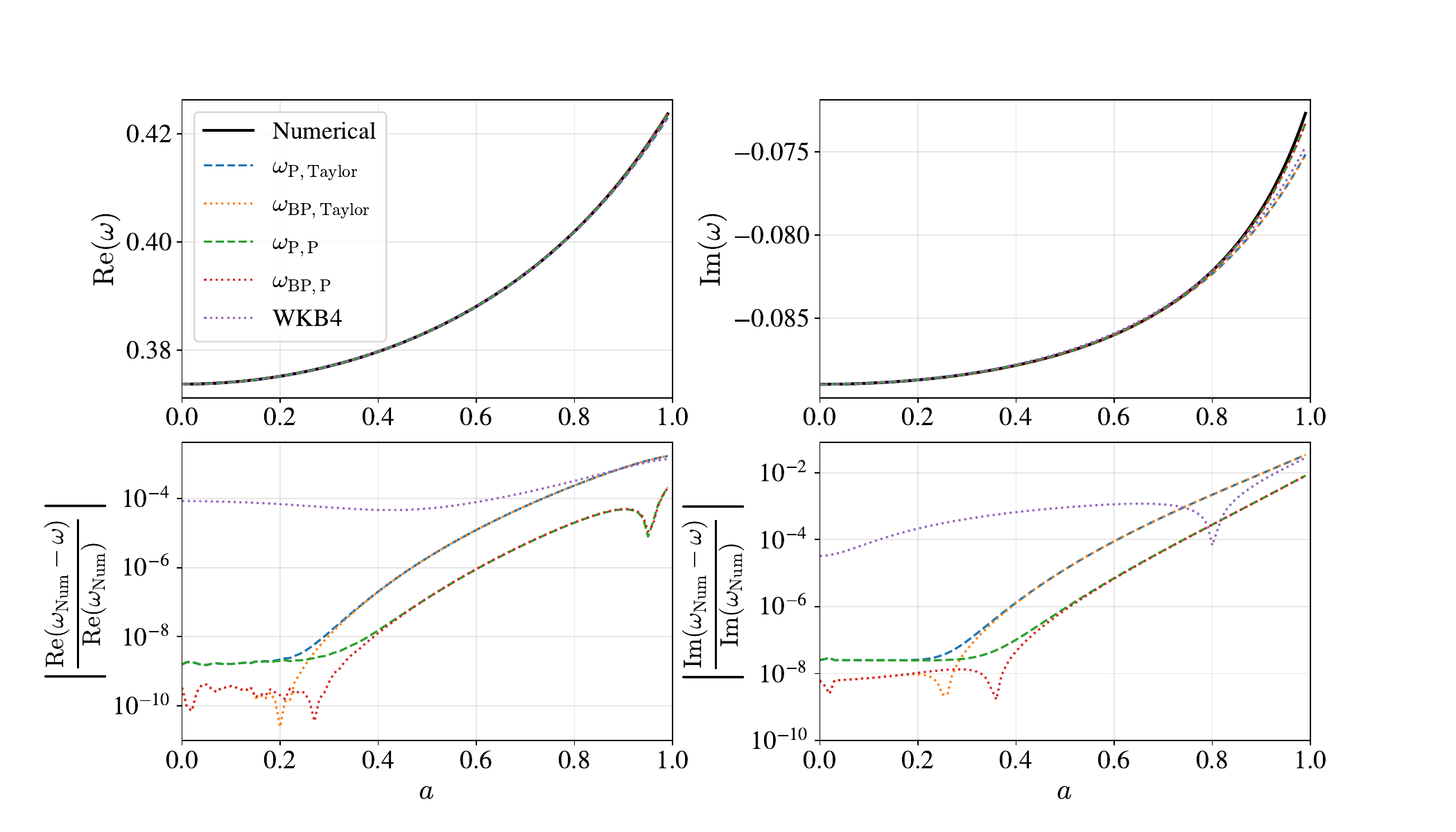}
    \caption{Semi-analytic resummed WKB results expanded to $a^8$ for the $(l,m,n)=(2,0,0)$ mode, compared to numerical results from Leaver's method. The dashed blue and dotted orange curves show $\omega_{\rm P,Taylor}$ and $\omega_{\rm BP,Taylor}$, respectively. The dashed green and dotted red curves show $\omega_{\rm P,P}$ and $\omega_{\rm BP,P}$, respectively.
    Purple lines show frequencies at the 4th order WKB, which is the optimal asymptotic order of the WKB approximation. The range of $a$ is $[0,0.99]$.  }
    \label{fig:200mode_slow}
\end{figure*}

\section{Accuracy and breakdown of resummed Kerr WKB frequencies} \label{sec: QNM frequencies}

This section tests the two constructions of the previous section against numerical QNM frequencies from Leaver's method. We first present the 21st-order slow-rotation results and their range of validity. We then use the 41st-order fixed-spin calculation to study larger spins, and show that the resummed WKB approximation remains accurate for damped modes but breaks down near the zero-damped-mode branch.

\subsection{Slow-rotation frequencies from 21st-order resummation} \label{sec: slow-rotation frequencies}

Let us focus on the results of our resummed WKB approach of Sec.~\ref{sec: slow-rotation} at 21st WKB order.
Because the $-m$ (retrograde) modes are excited in the opposite azimuthal direction to the $+m$ (prograde) modes, QNM frequencies for $(l,m,n)=(2,-1,0)$ or $(l,m,n)=(2,-2,0)$ modes can be obtained by replacing $a$ with $-a$ in the expressions for $(l,m,n)=(2,1,0)$ or $(l,m,n)=(2,2,0)$ modes, respectively.
Therefore, we present our solutions for the $(l=2,m,n=0)$ fundamental modes with nonnegative $m$, namely 
\bw
\begin{equation}
\begin{split}
    \omega_{\rm P,Taylor}^{(2,0,0)} &= (0.37367168 - 
   0.08896232i) + (0.03591326 + 
    0.00638179 i) a^2  + (0.00968815+ 
    0.00405142 i) a^4 \\& + (0.00359183 + 
    0.00240858 i) a^6  + (0.00152089 + 
    0.00149520 i) a^8 +\mathcal{O}(a^{10}) \label{eq: 200 pade slow}
\end{split}
\end{equation}

\begin{equation}
\begin{split}
    \omega_{\rm BP,Taylor}^{(2,0,0)} &= (0.37367168 - 
   0.08896232i) + (0.03591313 + 
    0.00638180 i) a^2 + (0.00968815+ 
    0.00405144 i) a^4 \\& + (0.00359183 + 
    0.00240859 i) a^6 + (0.00152067 + 
    0.00149581 i) a^8 +\mathcal{O}(a^{10}) \label{eq: 200 borel-pade slow}
\end{split}
\end{equation}

\begin{equation}
\begin{split}
    \omega_{\rm P,Taylor}^{(2,1,0)} &= (0.37367168 - 0.08896232i) + (0.06288308 + 0.00099794i)a  + (0.04486993 + 0.00609056i)a^2 \\& + (0.02181646 + 0.00287917i)a^3  + (0.01628597 + 0.00407896i)a^4  + (0.01093698 + 0.00254802i)a^5 \\& + (0.00838318 + 0.00276022i)a^6+ (0.00643122 + 
     0.00203498 i) a^7 + (0.00509854 + 
    0.00223134 i) a^8+\mathcal{O}(a^{9})
\end{split}
\end{equation}

\begin{equation}
\begin{split}
    \omega_{\rm BP,Taylor}^{(2,1,0)} &= (0.37367168- 0.08896232i)  + (0.06288308+ 0.00099793i)a  + (0.04486972+ 0.00609026i)a^2 \\& + (0.02181684+ 0.00287907i)a^3 + (0.01628257+ 0.00408181i)a^4 + (0.01091889+ 0.00257293i)a^5 \\& + (0.00826275+ 0.00268155i)a^6+(0.00659954 + 
     0.00215955 i) a^7 + (0.00484878 - 
    0.00003082 i) a^8+\mathcal{O}(a^{9})
\end{split}
\end{equation}

\begin{equation}
\begin{split}
    \omega_{\rm P,Taylor}^{(2,2,0)} &= (0.37367168 - 
   0.08896232i) + (0.12576617 + 
    0.00199587i)a  + (0.07174034 + 
    0.00521687i)a^2 \\& + (0.04802482 + 
    0.00440850i)a^3  + (0.03514272 + 
    0.00408180i)a^4  + (0.02707181 + 
    0.00382616i)a^5 \\& + (0.02198365 + 
    0.00340692i)a^6+(0.01664090 - 
     0.00227951 i) a^7 + (0.01754884 + 
    0.05968308 i) a^8+\mathcal{O}(a^{9})
\end{split}
\end{equation}

\begin{equation}
\begin{split}
    \omega_{\rm BP,Taylor}^{(2,2,0)} &= (0.37367168 - 
   0.08896232i) + (0.12576616 + 
    0.00199587i)a  + (0.07173949+ 
    0.00521563i)a^2 \\& + (0.04802811+ 
    0.00440832i)a^3 + (0.03508335+ 
    0.00412647i)a^4 + (0.02715913+ 
    0.00382530i)a^5 \\& + (0.02207878+ 
    0.00364686i)a^6 + (0.01655330 + 
     0.00151280 i) a^7 + (0.00652661 + 
    0.01125268 i) a^8+\mathcal{O}(a^{9})
\end{split}
\end{equation}
\ew
%
\begin{figure*}[ht!]
    \centering
    \includegraphics[clip=true,width=\textwidth]{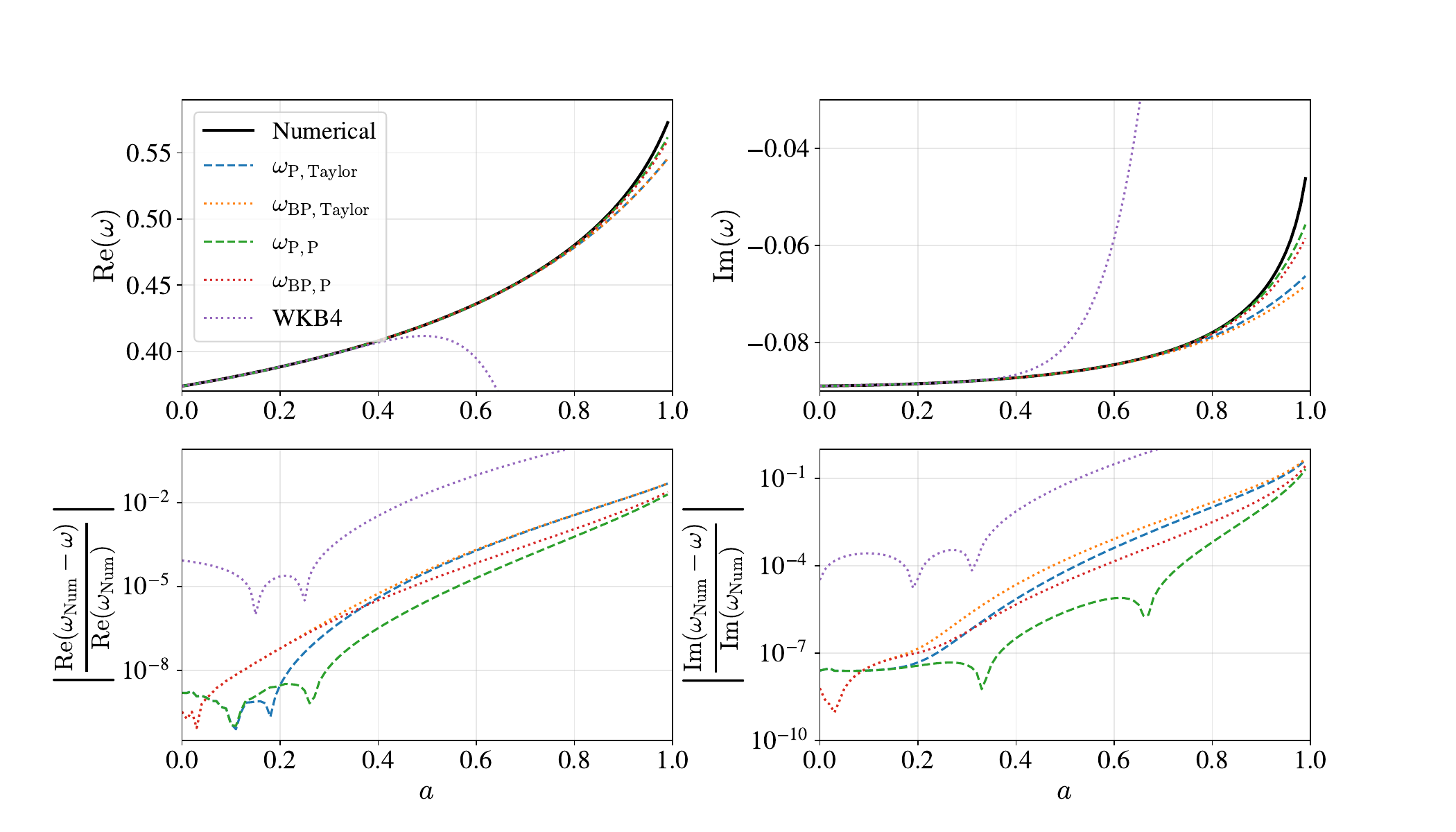}
    \\
    \vspace{-0.25cm}
    \includegraphics[clip=true,width=\textwidth]{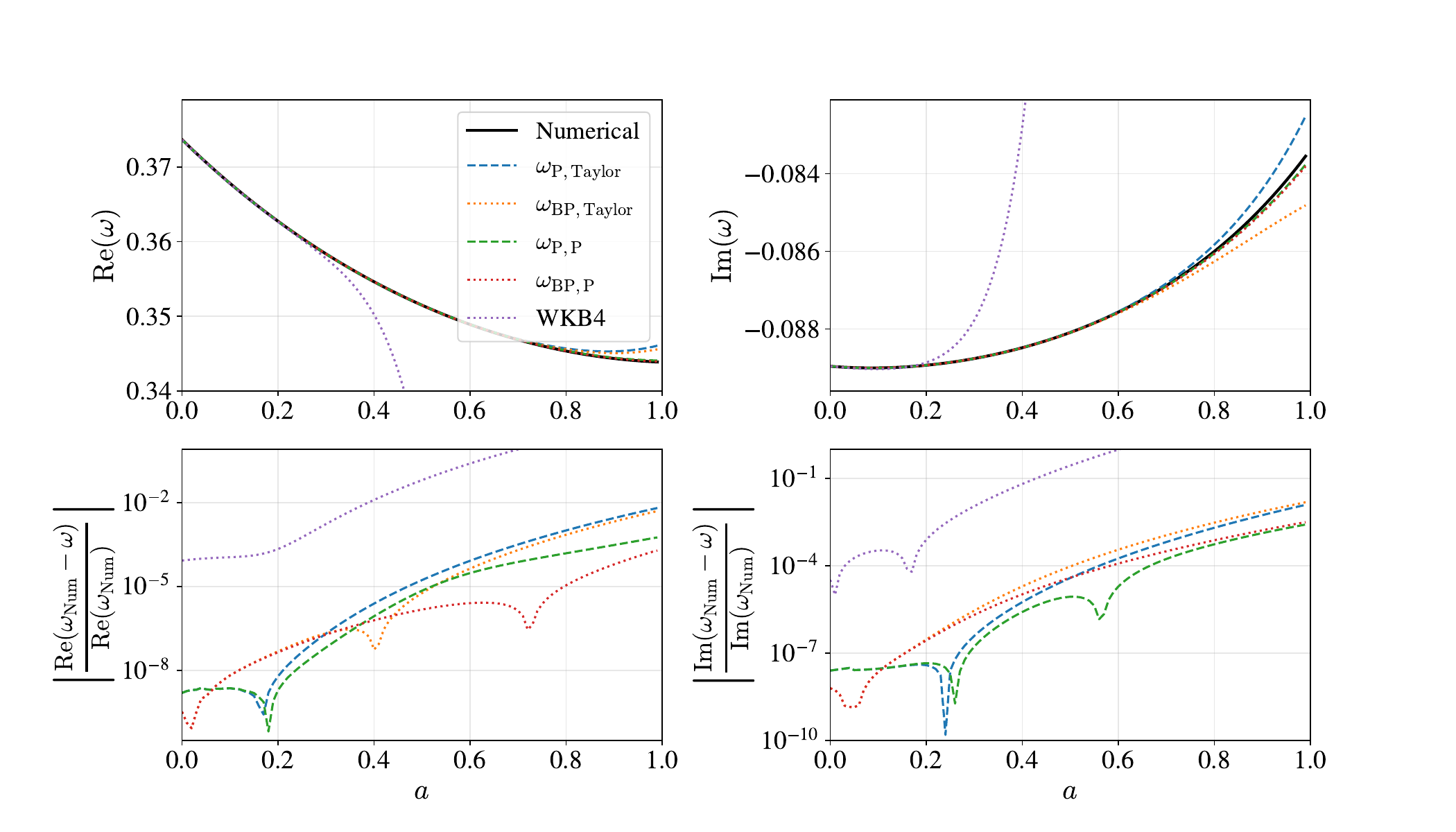}
    \caption{Semi-analytic, resummed WKB results expanded to $a^8$, compared to numerical results from Leaver's method for the $(l,m,n)=(2,1,0)$ mode (top) and the $(l,m,n)=(2,-1,0)$ mode (bottom). Label conventions are the same as those of Fig.~\ref{fig:200mode_slow}, with $a \in [0,0.99]$.  }\label{fig:2m0mode_slow}
\end{figure*}
\begin{figure*}[ht!]
    \centering
    \includegraphics[clip=true,width=\textwidth]{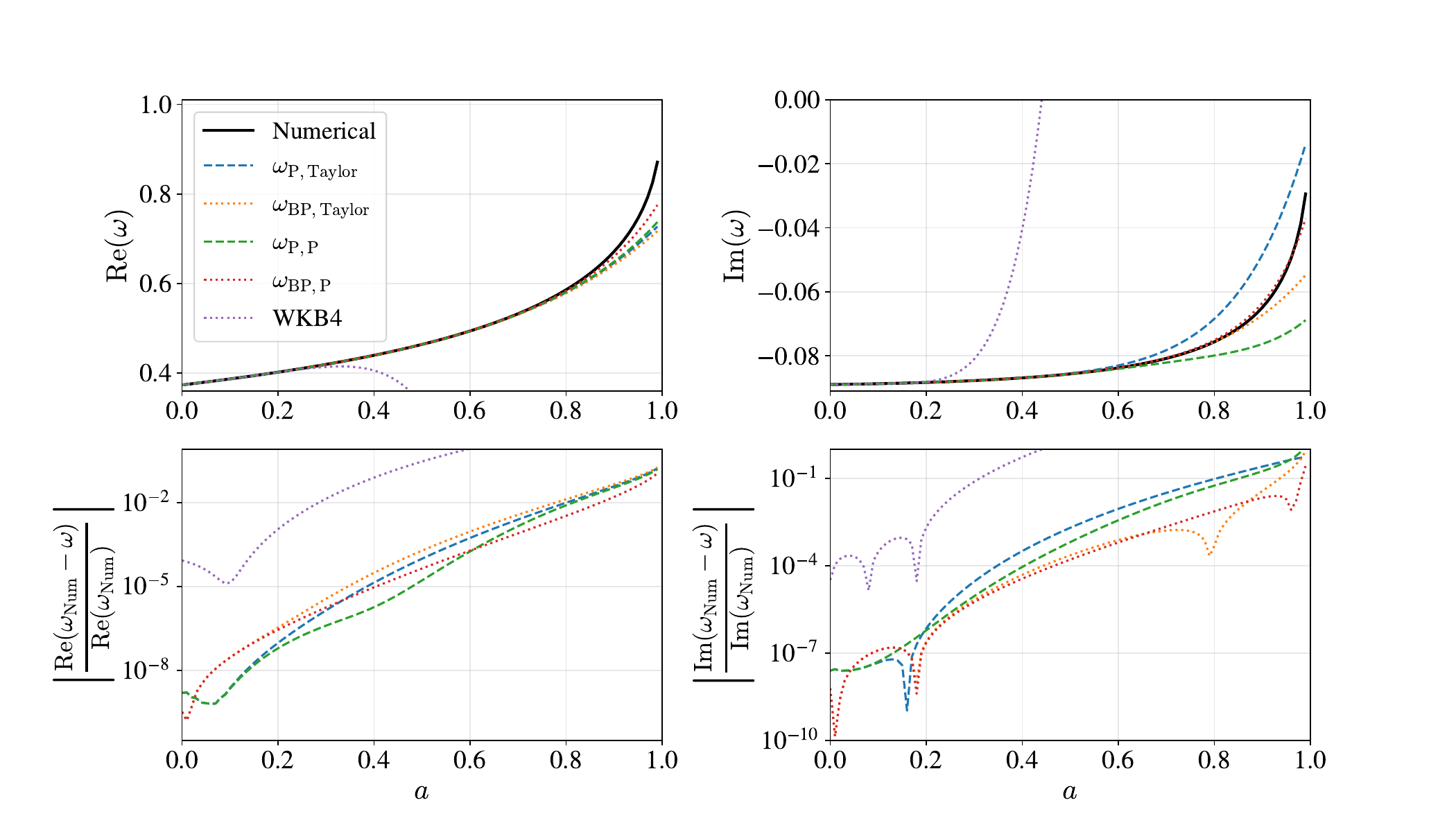}
    \\
    \vspace{-0.25cm}
    \includegraphics[clip=true,width=\textwidth]{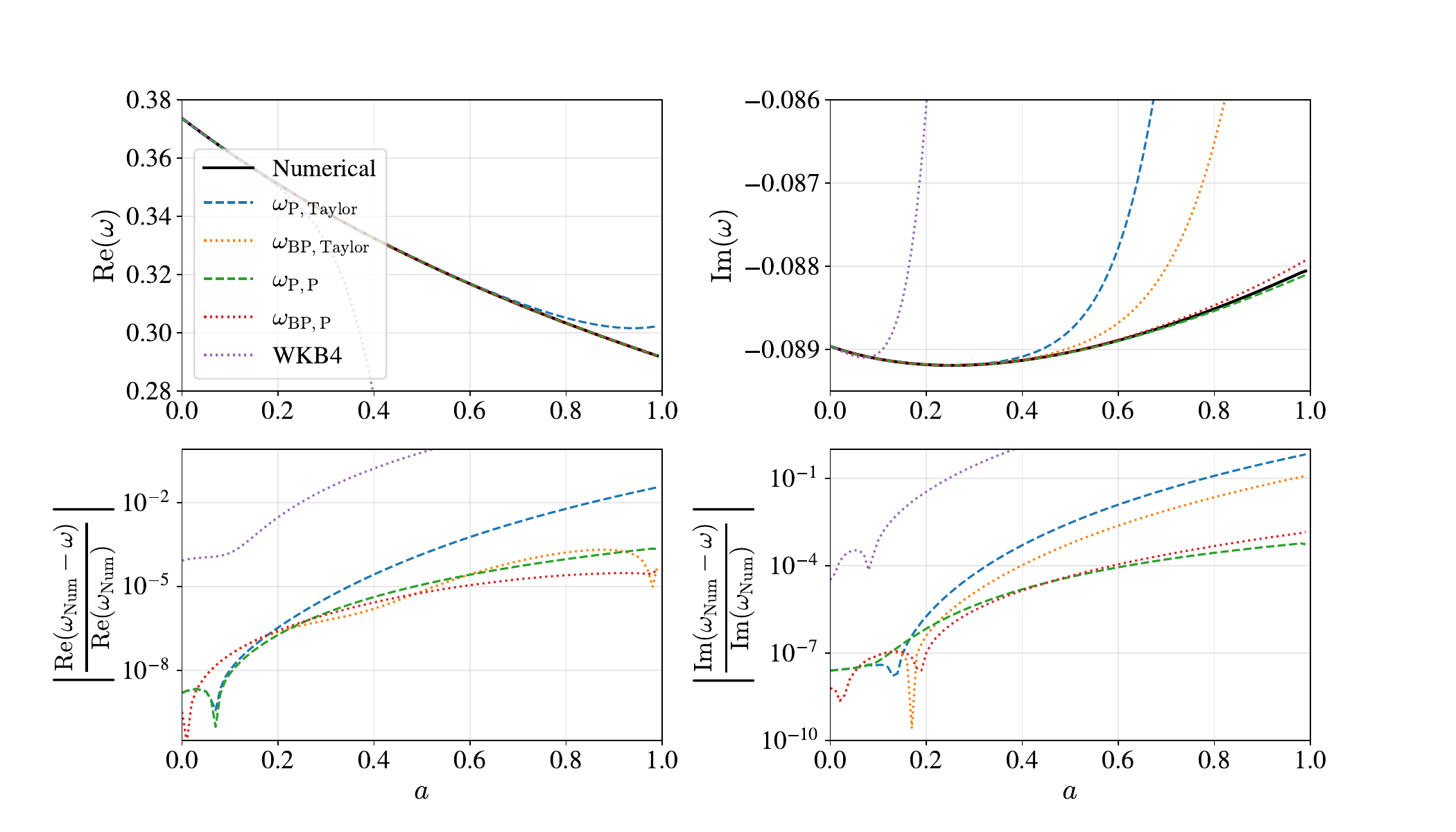}
    \caption{Same as Fig.~\ref{fig:2m0mode_slow} but for the $(l,m,n)=(2,2,0)$ mode (top) and the $(l,m,n)=(2,-2,0)$ mode (bottom). Label conventions are the same as those of Fig.~\ref{fig:200mode_slow}, with $a \in [0,0.99]$.  }\label{fig:2m2mode_slow}
\end{figure*}

The expressions above provide good approximations to the QNM frequencies in the slow-rotation regime.

Comparisons between our semi-analytic $(l,m,n)=(2,0,0)$ mode frequencies up to $a^8$ and numerical results~\cite{Cook:2014cta,Motohashi:2024fwt,Stein:2019mop} from Leaver's method are plotted in Fig.~\ref{fig:200mode_slow}. 
To further improve our approximations, we applied the second Pad\'e resummation over $a$ (see Eq.~(\ref{eq: additional pade})) to both Eq.~(\ref{eq: 200 pade slow}) and Eq.~(\ref{eq: 200 borel-pade slow}).
As shown in Fig.~\ref{fig:200mode_slow}, this second Pad\'e resummation improves the fractional difference approximately by an order of magnitude when $a>0.4$. Therefore, we will apply the same second Pad\'e resummation for all other modes in this subsection.

To benchmark the resummed results against ordinary WKB expansions, we also show in Fig.~\ref{fig:200mode_slow} the fourth-order WKB frequency without any Pad\'e or Borel-Pad\'e resummation, expanded consistently through $a^8$. We do not plot higher-order unresummed WKB curves here, because Fig.~\ref{fig:Sch WKB l2} already shows the relevant point: after the optimal asymptotic order, adding further WKB terms makes the un-resummed approximation worse rather than better. For $a<0.4$, where the slow-rotation expansion is most reliable, the resummed WKB results are more accurate than the ordinary fourth-order WKB approximation. At larger spins, the slow-rotation expansion itself begins to break down, so all semi-analytic, slow-rotation approximations accumulate larger errors. Even in this regime, however, the higher-order resummed curves remain closer to Leaver's results than the ordinary WKB curve. For rapidly rotating black holes, we  switch to the fixed-spin iterative method of Sec.~\ref{sec: iterative WKB}.

 We have also compared our fourth-order WKB frequencies in the slow-rotation limit with those recently reported by Tang et al.~\cite{Tang:2025qaq} for the same Kerr modes and find quantitative agreements (relative difference of the complex frequency magnitude around $10^{-4}$ for $n=0$ modes).
 Thus, the comparison made here should not be interpreted as a disagreement with their low-order WKB calculation. 
 Rather, our point is that the ordinary WKB series is asymptotic, so increasing the WKB order without resummation does not, in general, yield a systematically improving approximation.

\begin{figure*}[!ht]
    \centering
    \includegraphics[clip=true,width=\textwidth]{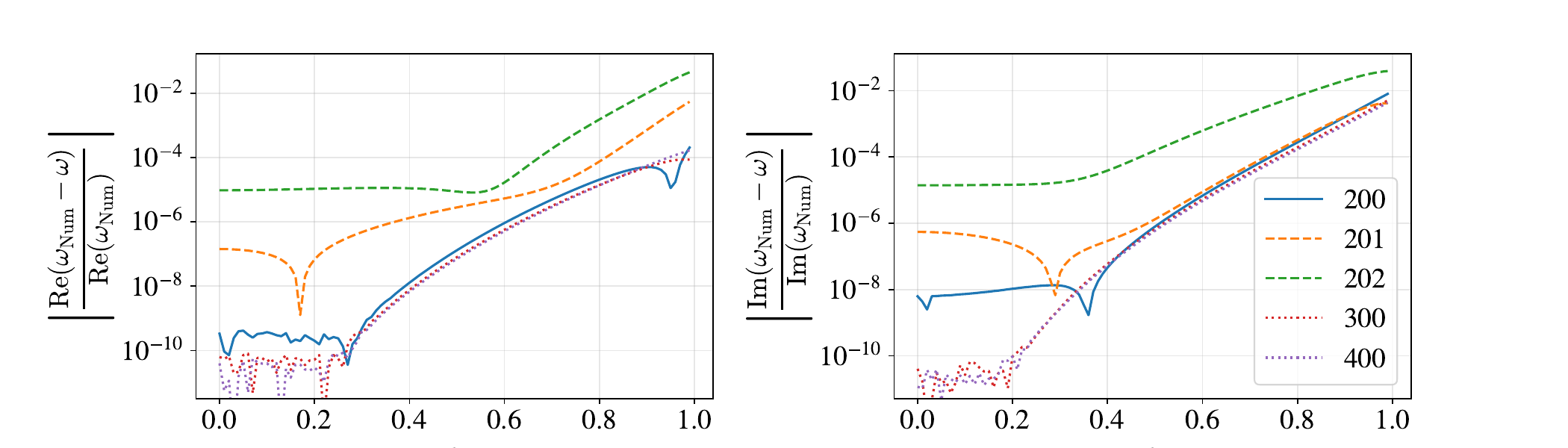}
    \caption{Fractional differences between the semi-analytic, double-resummed WKB frequencies and numerical results from Leaver's method for different overtone numbers $n$ and angular quantum numbers $l$, expanded to $a^8$. The labels denote the $(l,m,n)$ mode. The fractional difference decreases as $l$ increases or $n$ decreases.}
    \label{fig:diff_n&l}
\end{figure*}
\begin{figure*}[!ht]
    \centering
    \includegraphics[clip=true,width=\textwidth]{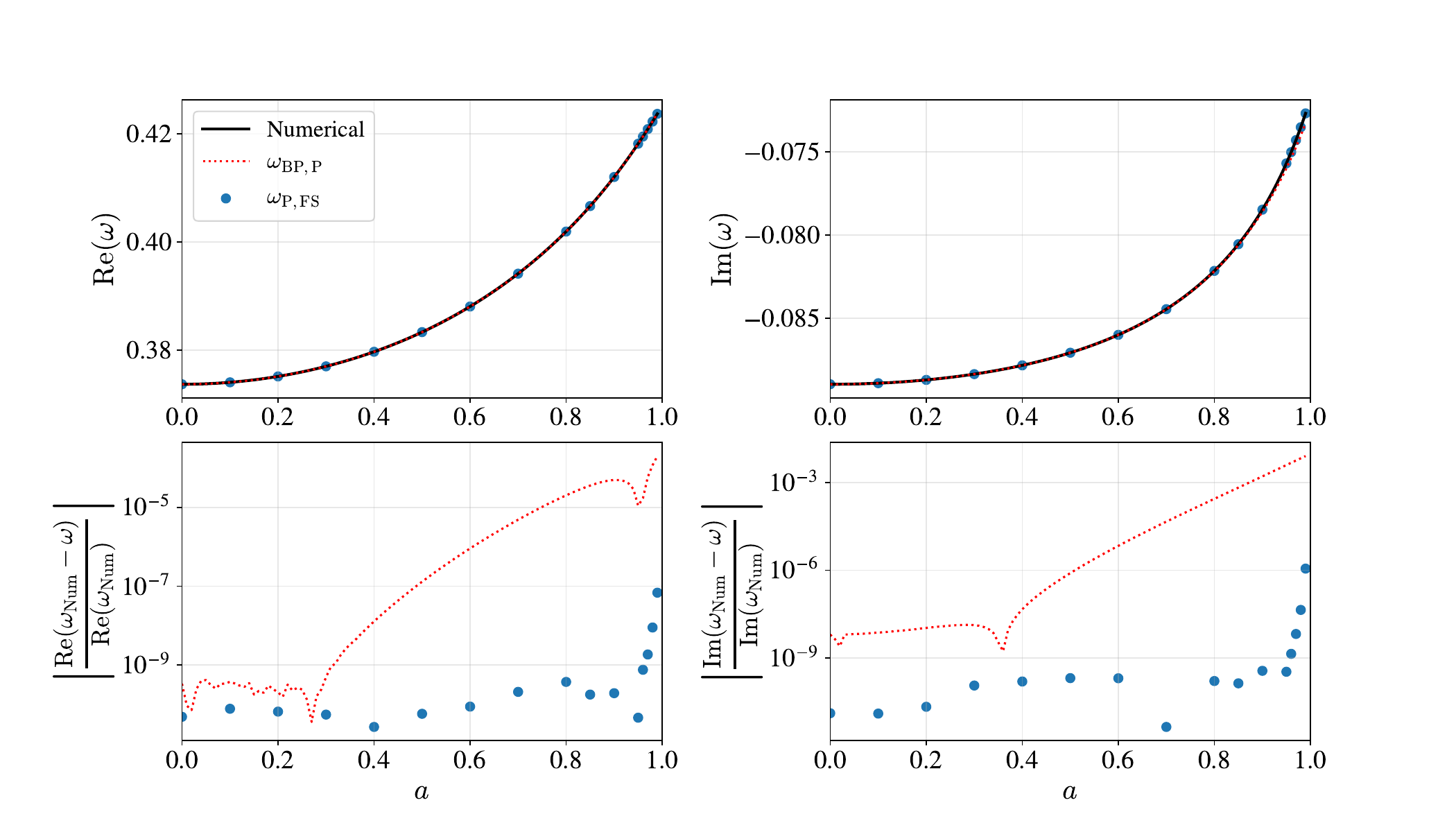}
    \caption{Comparison of the QNM frequencies between $\omega_{\rm P,FS}$ and numerical values obtained from Leaver's method for the $(l,m,n) = (2,0,0)$ mode. The blue dots are frequencies obtained by iterating the resummed 41st-order WKB approximation with $a$ values chosen between $0$ and $0.99$. The black solid curve is the numerical frequencies from Leaver's method. 
    The red dotted curves show the slow-rotation result $\omega_{\rm BP,P}$, included for reference. }
    \label{fig:200Numerical}
\end{figure*}

\begin{figure*}[!ht]
    \centering
    \includegraphics[width=\linewidth]{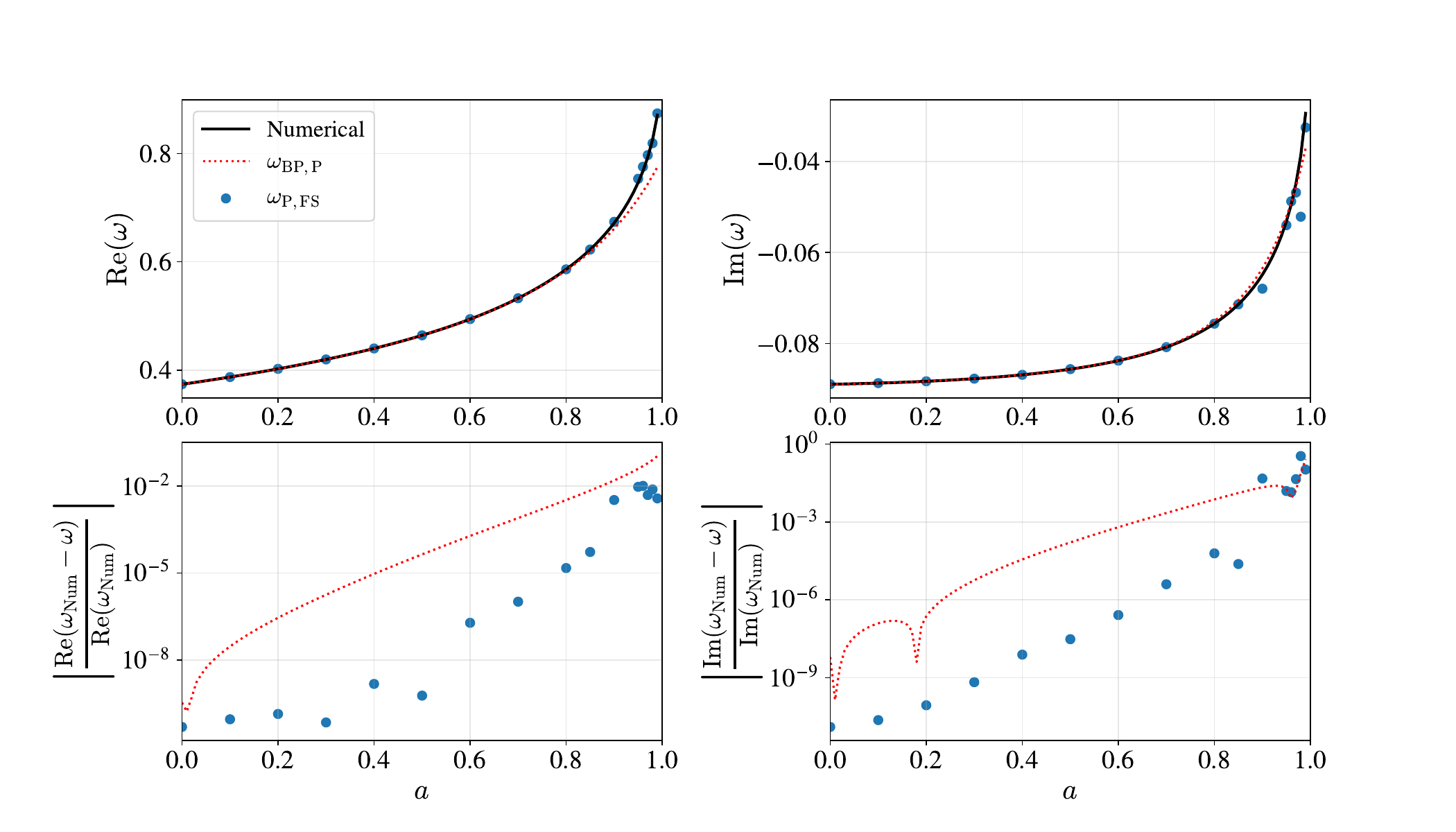}
    \caption{Comparison of the QNM frequencies between iterative resummed WKB and numerical values obtained from Leaver's method for the $(l,m,n) = (2,2,0)$ mode. Label conventions are the same as Fig.~\ref{fig:200Numerical}. }
    \label{fig:220Numerical}
\end{figure*}

We present our semi-analytic results for $(l,m,n)=(2,\pm1,0)$ and $(2,\pm2,0)$ modes in Figs.~\ref{fig:2m0mode_slow} and~\ref{fig:2m2mode_slow}. 
The semi-analytic frequencies obtained via resumming the WKB series at the 21st order and expanded up to $a^8$, and the 4th order WKB approximation (expanded to $a^8$) are plotted in purple dotted lines for comparison.
Comparing the fractional differences, we find that the double-resummed curves, in which the WKB series and the spin series are both resummed, give the best overall agreement with the numerical frequencies.
Even though these two methods are of similar accuracy, the $\omega_{\rm BP,P}$ result is slightly better than the $\omega_{\rm P,P}$ result for $(l,m,n)=(2,\pm2,0)$ modes in the $a>0.4$ region. 
For simplicity, the $\omega_{\rm BP,P}$ result will represent the "semi-analytic method" in the rest of the subsection.

We also find that the slow-rotation expansion is more accurate for retrograde modes rather than prograde modes. 
For example, $(l,m,n)=(2,2,0)$ and $(l,m,n)=(2,1,0)$ modes have fractional error greater than $10^{-3}$ for the real part of the frequency when $a>0.8$, while all other fundamental modes are all below $10^{-3}$.
This is primarily because of the drastic change in QNM frequencies for prograde modes at large spins, especially the $(l,m,n)=(2,2,0)$ mode, which evolves into zero damped modes in the extremal limit.
Due to the completely different origins of the zero-damped modes and damped modes, it might be impractical to approximate the whole QNM spectrum by expanding around $a=0$.

The QNM frequencies for different overtone numbers $n$ and different angular quantum numbers $l$ are plotted in Fig.~\ref{fig:diff_n&l}.
The fractional difference in Fig.~\ref{fig:diff_n&l} decreases as $l$ increases or $n$ decreases, which can be explained by the nature of the WKB approximation.
First, because the wave function of the higher overtones is more widespread compared to the base $n=0$ mode, the WKB approximation is expected to be less accurate for higher overtones, since the WKB approach expands around a narrow region near the potential peak. 
Second, the effective potential becomes smoother as $l$ increases, thereby improving the accuracy of the WKB approximation. 
Additionally, it can be shown that the fractional difference between the leading order WKB approach and numerical values scales as $1/l^2$ in the eikonal limit \cite{Yang:2012he}. 
The semianalytic QNM frequencies of the modes in Fig.~\ref{fig:diff_n&l} are provided in the supplemental Mathematica notebook.

\subsection{Fixed-spin frequencies and the onset of zero-damped-mode breakdown} \label{sec: iterative WKB frequencies}

In the large-spin regime, the modes relevant for our comparison can approach either the damped-mode branch or the zero-damped-mode branch.
Here we focus on two typical modes:
\begin{itemize}
    \item the $(l,m,n)=(2,0,0)$ mode, which evolves into damped modes in the extremal limit,
    \item the $(l,m,n)=(2,2,0)$ mode, which has a zero-damped-mode limit.
\end{itemize}

The $\omega_{\rm P,FS}$ results at 41st WKB order are plotted as blue dots in Figs.~\ref{fig:200Numerical} and~\ref{fig:220Numerical}, in the spin range $a\in[0,0.99]$.
For comparison, the slow-rotation expansion results are plotted as red dotted curves.
For the $(2,0,0)$ mode, the fractional difference for the real part of $\omega_{\rm P,FS}$ remains below $10^{-7}$ for $a\leq0.99$, which indicates the resummed WKB approach works well for modes with a damped-mode limit. 
However, the fractional difference for the real part of $\omega_{\rm P,FS}$ is above $10^{-3}$ for $(2,2,0)$ when $a>0.9$. 

\begin{figure*}[!htbp]
    \centering
    \includegraphics[width=0.49\textwidth]{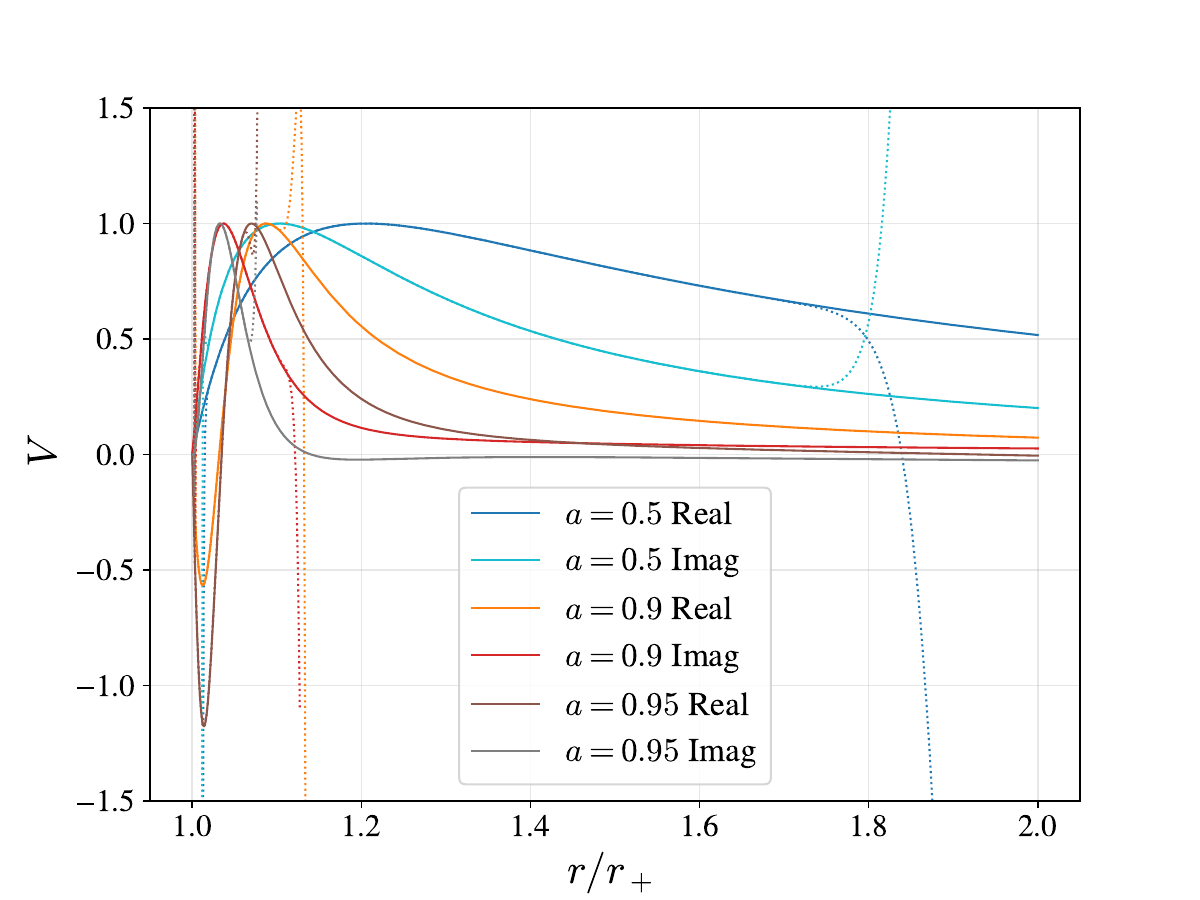}
    \includegraphics[width=0.49\textwidth]{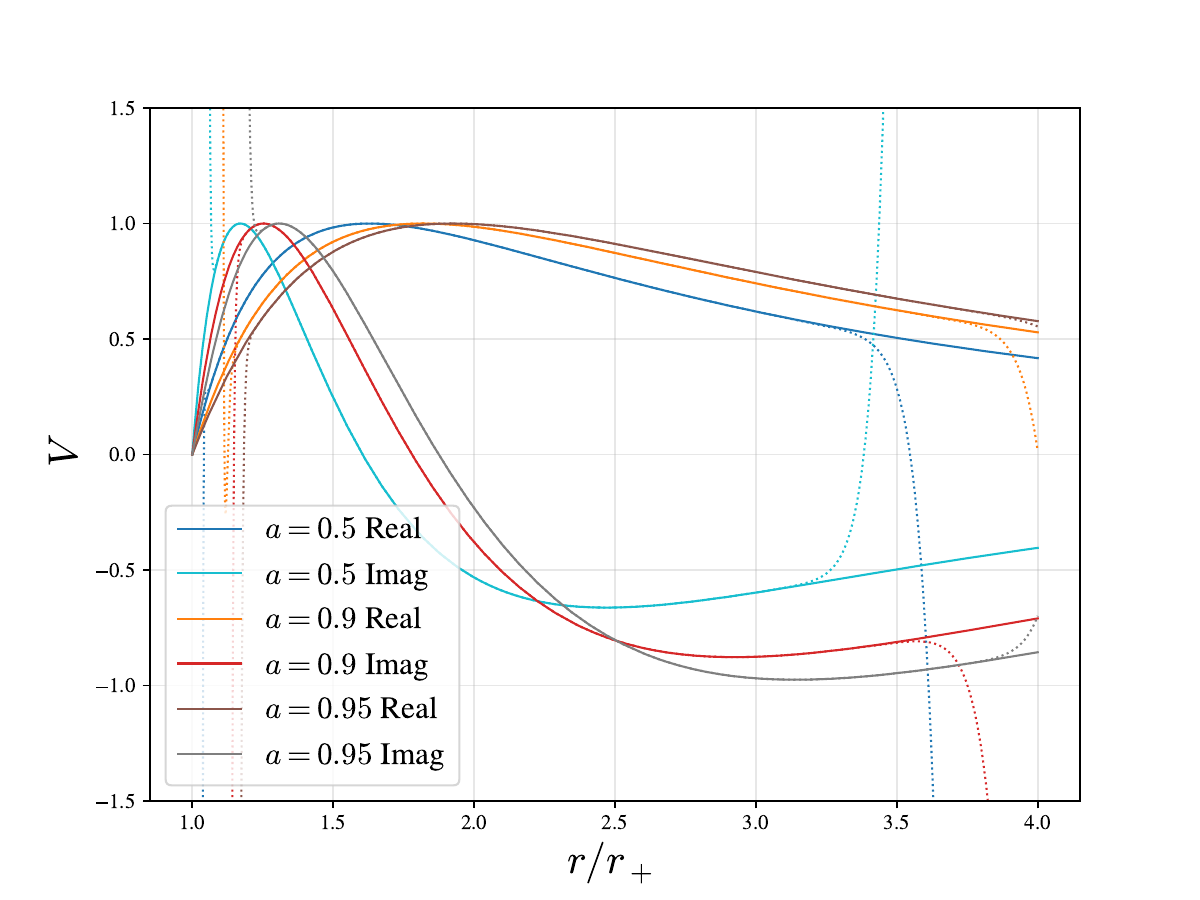}
    \caption{The Taylor-expanded (in $r_\star$) potential used in the WKB approximation compared to the full potential for different dimensionless spins $a$, focusing on the $(l,m,n)=(2,2,0)$ mode (left) and the $(l,m,n)=(2,0,0)$ mode (right). The potentials are expanded around $r_0$, where $\frac{d V}{d r}|_{r=r_0}=0$. The solid curves are the Chandrasekhar-Detweiler potential, and the dotted curves are the 82nd-order Taylor-expanded (in $r_\star$) potential, corresponding to the 41st-order WKB approximation. }\label{fig:Taylor potential}
\end{figure*}

The physical reason for the large fractional difference near the zero-damped-mode limit can be explained with Fig.~\ref{fig:Taylor potential}, where the Taylor-expanded potentials used for the WKB analysis are plotted in dotted curves, and the full potentials are plotted in solid curves. 
The real and imaginary parts of the potentials are normalized so that the maximum potential is $1$.
The radial coordinate is normalized by the outer horizon radius $r_+$.
The potential of the $(2,0,0)$ mode remains well approximated by the Taylor series in the region $r/r_+\in(1.3,3.0)$, and this well-approximated region widens as $a$ increases. 
However, the potential of the $(2,2,0)$ mode varies rapidly near the horizon as $a$ approaches extremality, causing the well-approximated region to shrink rapidly as $a$ increases. 
Since WKB approximations prefer smooth potentials, this rapid near-horizon variation reduces the accuracy of the WKB approximation near the zero-damped-mode limit. The cause of this near-horizon oscillation of the potential will be examined next.


\subsubsection{Near-horizon Poles of the Chandrasekhar-Detweiler Potential}

The rapid variation of the Chandrasekhar-Detweiler potential for modes near the zero-damped limit is due to the potential's poles near the outer horizon. 
The positive roots of $\Delta$ and $\bar{\rho}$ are $r_\pm=1\pm\sqrt{1-a^2}$ and $r_{\omega+} = \sqrt{am/\omega-a^2}$, respectively. 
For the modes with zero-damped limits, $r_+$ and $r_{\omega+}$ both approach unity as $a\rightarrow1$, but at slightly different speeds.
In the near-extremal limit, defining $\epsilon = 1-a$, the shift in the QNM frequency is proportional to $\sqrt{\epsilon}$~\cite{Hod:2008zz,Hod:2009td,Yang:2013uba}, so
\begin{equation}
    \omega \sim \frac{m}{2}-\delta'\sqrt{\epsilon},
\end{equation}
where $\delta'$ is a complex constant~\footnote{Here we have absorbed a $\sqrt{2}$ constant into $\delta'$.}~\cite{Hod:2008zz,Hod:2009td,Yang:2013uba}.
Hence, at leading order in $\epsilon$, we have $r_\pm = 1\pm \sqrt{2\epsilon}$ and $r_{\omega+}=1+ ({2 \delta'}/{m})\sqrt{\epsilon}$. 

In the near-extremal regime, the Chandrasekhar-Detweiler potential has poles near the outer horizon in the complex plane. For example, the first term of the potential, $({-\Delta^2}/{\bar{\rho}^8}) \beta_2$, has poles at $r=r_{\omega+}$, which cause rapid variation. However, this divergence is suppressed by $\Delta$ (and actually goes to zero) at the outer horizon. 

The same scaling analysis can be applied to the rest of the potential. In the near-horizon, near-extremal region, we take $\epsilon\ll1$ and $r-r_+=\mathcal{O}(\sqrt{\epsilon})$. Since $r_+-r_-=\mathcal{O}(\sqrt{\epsilon})$ and $r_{\omega+}-1=\mathcal{O}(\sqrt{\epsilon})$, the quantities $r-r_-$, $r-1$, and $r-r_{\omega+}$ are also $\mathcal{O}(\sqrt{\epsilon})$ in this region. With this scaling, the Chandrasekhar-Detweiler potential can be expressed as
\bw
\begin{equation}
\begin{split}
    V=&\frac{3}{16}\,\frac{(r-r_+)^2(r-r_-)^2}{(r-r_{\omega+} )^4}+\frac{\kappa (r-r_+)(r-r_-)}{12(r-1)(r-r_{\omega+})+4\lambda(r-r_{\omega+})^2-6(r-r_+)(r-r_-)}+\cdots  +\mathcal{O}(\sqrt{\epsilon}),
\end{split} \label{eq: NHEK CD potential}
\end{equation}
\ew
where $\kappa$ and $\lambda$ are evaluated at $a=1$.   
The third term at $\mathcal{O}(1)$ is omitted here because it is algebraically lengthy; its expression is given in Appendix~\ref{app: CD potential}.
All three terms have poles near the outer horizon in the complex $r$ plane, and the poles are suppressed by $\Delta$ at $r_+$. 

Therefore, in the extremal limit, discarding the $\mathcal{O}(\sqrt{\epsilon})$ term, one may examine the near-horizon and near-extremal ($\sqrt{\epsilon}\ll1$) behaviour of the potential is
\begin{equation}
V=
\begin{cases}
\mathcal{O}(1),  & \text{if $(r-r_+)\sim \sqrt{\epsilon}$}, \\
V_R=\frac{9-3m^2+4 \,{_sA}_{lm}}{16},& \text{if $(r-r_+)/\sqrt{\epsilon}\gg 1 $},\\
V_L=0, & \text{if $(r-r_+)=0$},
\end{cases}  \label{eq: near NHEK potential}
\end{equation}
This indicates that the near-horizon region has an asymptotic structure distinct from the outer-region potential, so the Taylor expansion about the peak cannot uniformly approximate both regions as extremality is approached.
The left boundary of the region is located at $r-r_+=0$ where $V=V_L=0$. The right asymptotic boundary is at $(r-r_+)/\sqrt{\epsilon}\gg1$.
At the right boundary, the potential asymptotically approaches $V=V_R=({9-3m^2+4 \,{_sA}_{lm}})/{16}$. 

\subsubsection{Connection to the NHEK Boundary Equation}

The boundary behavior of this near-horizon oscillating region can be naturally compared with the throat region of extremal Kerr. In particular, the right-boundary constant $V_R$ matches the boundary form of the radial equation for gravitational perturbations in the NHEK geometry, as we now show.

In the NHEK geometry, the radial master equation for gravitational perturbation is~\cite{Dias:2009ex}
\begin{equation}
    \frac{d }{dy}\left[(1+y^2)\frac{d}{dy}\Phi\right]-\left[\mu^2-\frac{(\omega+q y)^2}{1+y^2}\right]\Phi = 0, \label{eq: NHEK radial}
\end{equation}
with $y=[{(1+\tilde{t}^2)\tilde{r}^2-1}]/({2\tilde{r}})$. 
Here, $\tilde{t}$ and $\tilde{r}$ arise from the NHEK limit~\cite{Bardeen:1999px}
\begin{equation}
    \tilde{t} = \lim_{\tilde{\lambda}\to 0} \tilde{\lambda}\, t \,\big|_{a=1},
    \qquad
    \tilde{r} = \lim_{\tilde{\lambda}\to 0} \frac{r-M}{\tilde{\lambda}}\,\bigg|_{a=1},
\end{equation}
where $\tilde{\lambda}$ is the scaling factor.
The remaining constants are $q = m - si$ and $\mu^2 = {}_{s}A_{lm} + s - 2ism - {3m^2}/{4}$, with $s=-2$ for the gravitational perturbations considered here.

At the boundary of the NHEK geometry, where $y=\pm \infty$, Eq.~(\ref{eq: NHEK radial}) becomes
\begin{equation}
    \frac{d }{dy}[y^2\frac{d}{dy}\Phi]-(\mu^2-q^2)\Phi = 0, \label{eq: NHEK radial infinity}
\end{equation}
and the wavefunction becomes $\Phi=R/y^2$~\cite{Dias:2009ex}. By Eq.~(\ref{eq:Z-R relation}), we find that 
\begin{equation}
\lim\limits_{y\rightarrow\pm \infty} Z\propto \frac{R}{y^{3/2}},    
\end{equation} 
so $\Phi\propto Z/ \sqrt{y}$. 

Rewriting Eq.~(\ref{eq: NHEK radial infinity}) in terms of Z, the radial master equation for the gravitational master function at the boundary becomes
\begin{equation}
    \frac{y}{2} \frac{d}{dy} \left(\frac{y}{2} \frac{d Z}{dy}\right)+\frac{9-3m^2+4 \,{_sA}_{lm}}{16} Z=0.
\end{equation}
The limit $\lim\limits_{y\rightarrow\pm \infty} {d}/{dr_\star}= ({y}/{2}) {d}/{dy}$. Hence, the boundary form of the NHEK radial equation contains the same constant $V_R=(9-3m^2+4\,{_sA}_{lm})/16$ that appears at the right boundary of the near-horizon Chandrasekhar-Detweiler potential.
For near-extremal Kerr black holes, when $\sqrt{\epsilon}$ and $\tilde{\lambda}$ scale together under a fixed temperature $T={\sqrt{\epsilon}}/({4\pi \tilde{\lambda}})$, the near-horizon geometry is also diffeomorphic to the NHEK throat geometry~\cite{Amsel:2009ev}.
Therefore, in the near-extremal limit, the throat scaling region has width $r-r_+=\mathcal{O}(\sqrt{\epsilon})$, which can be mapped to the near-horizon and near-extremal potential in Eq.~(\ref{eq: near NHEK potential}).

This scaling also explains the behavior seen in the left panel of Fig.~\ref{fig:Taylor potential}. As $\epsilon\ll1$, the region over which the Taylor-expanded potential remains accurate shrinks with the same near-horizon scale, $r-r_+=\mathcal{O}(\sqrt{\epsilon})$. The WKB approximation used here is local because it expands the potential about its peak and assumes that this local expansion captures the relevant barrier region. Near the zero-damped-mode limit, however, the potential develops structure on the throat scale, and this structure is not captured uniformly by the Taylor series about the peak. A more complete analysis would require mapping the oscillatory region of the Chandrasekhar-Detweiler potential to the finite-$y$ region of the NHEK geometry, which we leave to future work. As a consistency check, we show in Appendix~\ref{app: eikonal} that, in the combined eikonal and near-extremal limit, the same Chandrasekhar-Detweiler potential still reproduces the leading-order WKB result of Refs.~\cite{Yang:2012he,Yang:2013uba}.

\section{Conclusions and outlook}\label{sec: conclusions}

In this paper, we studied the high-order WKB expansion for Kerr QNM frequencies as an asymptotic series, and asked whether its high-order information can be made predictive by resummation. We constructed two implementations based on the Chandrasekhar-Detweiler potential. In the slow-rotation implementation, we expanded the potential, angular separation constant, peak location, and frequency in the spin, computed the WKB series through 21st order, and resummed it with Pad\'e and Borel-Pad\'e approximants. We then applied a second Pad\'e resummation to the spin expansion itself. This double-resummed slow-rotation calculation improves substantially over the ordinary fourth-order WKB approximation in its regime of validity, with the best agreement coming from the resummed spin series.

To go beyond the slow-rotation expansion, we also developed a fixed-spin Pad\'e-WKB method, in which the spin is fixed and the implicit Pad\'e-resummed WKB frequency equation is solved numerically. This calculation was carried out through 41st WKB order. For damped-mode branches, the method can be highly accurate. For example, for the $(l,m,n)=(2,0,0)$ mode, the fractional error in the real part remains below $10^{-7}$ relative to Leaver's method through $a=0.99$. The same is not true, however, for modes approaching the zero-damped-mode branch. For the $(2,2,0)$ mode, the error exceeds $10^{-3}$ for $a>0.9$. 

We traced this breakdown of the WKB approximation to the near-horizon, near-extremal structure of the Chandrasekhar-Detweiler potential. In this regime, poles of the potential approach the outer horizon on the same $\sqrt{1-a}$ scale that controls the zero-damped-mode frequency shift, and the local Taylor expansion about the potential peak no longer gives a uniform description of the relevant throat region. This same scale is also the one that appears in the near-horizon extremal Kerr geometry, and the boundary form of the near-horizon Chandrasekhar-Detweiler potential matches the boundary form of the NHEK radial equation.

Our results suggest a concrete route toward resummed WKB calculations of QNM frequencies beyond GR, provided the perturbation equations can be cast as a sufficiently smooth effective-potential problem. Recent work by Tang et al.~\cite{Tang:2025qaq} provides a first step in this direction, by applying low-order WKB methods to rotating black holes in GR and to selected perturbative beyond-GR settings. The natural next step is to combine this strategy with the modified Teukolsky formalism~\cite{Li:2022pcy,Wagle:2023fwl}, which is designed to describe perturbations of rotating black holes in more general theories without relying on a slow-rotation expansion. If the resulting radial equation can be transformed, in analogy with the Chandrasekhar-Detweiler transformation, into an equation with a suitable effective potential, then the resummed WKB methods developed here could be used to compute beyond-GR corrections to QNM frequencies at fixed spin.

The WKB treatment developed here may also be applied to coupled QNM systems, where the perturbations of the gravitational field talk to additional fields~\cite{Blazquez-Salcedo:2016enn,Cardoso:2009pk,Karikos:2026arz}. 
Coupled QNM systems require a matrix-valued potential rather than a single Chandrasekhar-Detweiler potential. 
In regimes where the system can be approximately diagonalized, or where the off-diagonal couplings can be treated perturbatively, WKB methods may still provide useful estimates for the eigenchannel frequencies~\cite{Hui:2022vov}.

Near extremality, beyond-GR frequency shifts may require additional care. In some quadratic gravity theories, these shifts can become large, and the perturbative expansion in the coupling may break down in the near-extremal limit~\cite{Husken:2026axq}. Our results point to a related difficulty: the rapid near-horizon variation of the potential may prevent a local high-order WKB expansion, resummed or otherwise, from remaining uniformly accurate. It would therefore be useful to develop a WKB expansion directly about the extremal limit. Such an expansion could then be matched to the slow-rotation result, extending the semi-analytic framework developed here across a larger portion of the Kerr parameter space.

\section{Acknowledgements}

J.H and N.Y. acknowledge support from the Simons Foundation through Award No. 896696,
 the Simons Foundation International through Award No. SFI-MPS-black hole-00012593-01,
 and the NSF through Grants No. PHY-25-12423.
 K.Y. acknowledges support from NSF Grant No. PHY-2309066 and PHYS-2339969.


\appendix

\section{Explicit Pad\'e resummation example}
\label{app:first-order-pade-example}

This appendix gives a concrete example of the notation introduced in Sec.~\ref{sec: slow-rotation}. We focus on the fundamental $(l,m,n)=(2,1,0)$ mode and keep only Pad\'e resummation of the WKB series. The Borel-Pad\'e case follows the same logic, with the first label changed from ${\rm P}$ to ${\rm BP}$. To keep the expressions short enough to display, we truncate the WKB series at $N_{\rm max}=6$ in this appendix. The main text uses the same procedure, but with the WKB series carried to higher order. The superscript $(6)$ below denotes $N_{\rm{max}}=6$ for the seventh-order WKB approximation.

At first order in the spin, the WKB coefficients may be written as
\begin{align}
    \epsilon_k(\omega,a)
    &=
    \epsilon_{k,0,0}
    +
    a
    \sum_{i=-1}^{1}
    \epsilon_{k,1,i}\omega^i
    +
    {\cal O}(a^2).
\end{align}
For each fixed pair $(j,i)$, define the seventh-order WKB coefficient series
\begin{equation}
    E^{(6)}_{j,i}(\hhbar)
    =
    \sum_{k=0}^{6}
    \epsilon_{k,j,i}\hhbar^k .
\end{equation}
For the mode used in this appendix, these coefficient series are
\begin{align}
    E^{(6)}_{0,0}
    &=
    0.035242726457
    -
    0.010056365000\,\hhbar
    \nonumber \\
    &+
    0.002003924238\,\hhbar^2
    -
    0.000247627939\,\hhbar^3
    \nonumber \\
    &-
    0.000007497493\,\hhbar^4
    +
    0.000028335933\,\hhbar^5
    \nonumber \\
    &
    -
    0.000038284491\,\hhbar^6,
    \\
    E^{(6)}_{1,-1}
    &=
    0.011110946982
    -
    0.004007722676\,\hhbar
    \nonumber \\
    &
    +
    0.001034008879\,\hhbar^2
    -
    0.000216609319\,\hhbar^3
    \nonumber \\
    &
    +
    0.000088362387\,\hhbar^4
    -
    0.000104484340\,\hhbar^5
    \nonumber \\
    &
    +
    0.000134500748\,\hhbar^6,
    \\
    E^{(6)}_{1,0}
    &=
    0,
    \\
    E^{(6)}_{1,1}
    &=
    -
    0.015052287878
    +
    0.000323598155\,\hhbar
    \nonumber \\
    &
    +
    0.000826480400\,\hhbar^2
    -
    0.000272899259\,\hhbar^3
    \nonumber \\
    &
    +
    0.000041547498\,\hhbar^4
    +
    0.000022751168\,\hhbar^5
    \nonumber \\
    &
    -
    0.000083824868\,\hhbar^6 .
\end{align}
The dimensionless value of the potential at the peak is
\begin{align}
    {\cal V}_0(\omega,a)
    &=
    0.151310605961
    \nonumber \\
    &
    +
    a
    \left(
    0.038007557913\,\omega^{-1}
    -
    0.120821975642\,\omega
    \right)
    \nonumber \\
    &
    +
    {\cal O}(a^2).
\end{align}

We now perform the first resummation. This is the Pad\'e resummation of the WKB series in $\hhbar$, not a resummation of the spin series. For a generic seventh-order series
\begin{equation}
    E^{(6)}(\hhbar)=\sum_{k=0}^{6}c_k\hhbar^k,
\end{equation}
the diagonal Pad\'e approximant is
\begin{equation}
    P^3_3\left[E^{(6)}\right]
    =
    \frac{
    p_0+p_1\hhbar+p_2\hhbar^2+p_3\hhbar^3
    }{
    1+q_1\hhbar+q_2\hhbar^2+q_3\hhbar^3
    }.
\end{equation}
The coefficients are fixed by requiring
\begin{equation}
    P^3_3\left[E^{(6)}\right]
    -
    E^{(6)}(\hhbar)
    =
    {\cal O}(\hhbar^7).
\end{equation}
Equivalently,
\begin{align}
    p_0 &= c_0,
    \\
    p_1 &= c_1+q_1c_0,
    \\
    p_2 &= c_2+q_1c_1+q_2c_0,
    \\
    p_3 &= c_3+q_1c_2+q_2c_1+q_3c_0,
\end{align}
where
\begin{equation}
\begin{pmatrix}
    c_3 & c_2 & c_1 \\
    c_4 & c_3 & c_2 \\
    c_5 & c_4 & c_3
\end{pmatrix}
\begin{pmatrix}
    q_1 \\ q_2 \\ q_3
\end{pmatrix}
=
-
\begin{pmatrix}
    c_4 \\ c_5 \\ c_6
\end{pmatrix}.
\end{equation}

Applying this construction to the coefficient series above gives
\begin{equation}
    P^3_3\left[E^{(6)}_{0,0}\right]
    =
    \frac{N_{0,0}}{D_{0,0}},
\end{equation}
with
\begin{align}
    N_{0,0}
    &=
    0.035242726457
    +
    0.065190298216\,\hhbar
    \nonumber \\
    &
    +
    0.002492801443\,\hhbar^2
    +
    0.000261538097\,\hhbar^3,
    \\
    D_{0,0}
    &=
    1
    +
    2.135097672088\,\hhbar
    +
    0.623112934604\,\hhbar^2
    \nonumber \\
    &
    +
    0.070847048965\,\hhbar^3.
\end{align}
Similarly,
\begin{equation}
    P^3_3\left[E^{(6)}_{1,-1}\right]
    =
    \frac{N_{1,-1}}{D_{1,-1}},
\end{equation}
where
\begin{align}
    N_{1,-1}
    &=
    0.011110946982
    +
    0.013820887774\,\hhbar
    \nonumber \\
    &
    +
    0.001153195272\,\hhbar^2
    +
    0.000051276242\,\hhbar^3,
    \\
    D_{1,-1}
    &=
    1
    +
    1.604598643024\,\hhbar
    +
    0.589506256390\,\hhbar^2
    \nonumber \\
    &
    +
    0.087417743023\,\hhbar^3.
\end{align}
Finally,
\begin{equation}
    P^3_3\left[E^{(6)}_{1,1}\right]
    =
    \frac{N_{1,1}}{D_{1,1}},
\end{equation}
where
\begin{align}
    N_{1,1}
    &=
    -
    0.015052287878
    -
    0.052099086593\,\hhbar
    \nonumber \\
    &
    -
    0.013761844688\,\hhbar^2
    +
    0.000803893480\,\hhbar^3,
    \\
    D_{1,1}
    &=
    1
    +
    3.482705431423\,\hhbar
    +
    1.044048736459\,\hhbar^2
    \nonumber \\
    &
    +
    0.142134358653\,\hhbar^3.
\end{align}

The WKB-Pad\'e-resummed object that enters the frequency equation is therefore
\begin{align}
    {\cal P}^{(6)}_{\rm P}(\omega,a;\hhbar)
    &=
    P^3_3\left[E^{(6)}_{0,0}\right]
    \nonumber \\
    &
    +
    a
    \left[
    \omega^{-1}
    P^3_3\left[E^{(6)}_{1,-1}\right]
    \right.
    \nonumber \\
    &
    \left.
    +
    \omega
    P^3_3\left[E^{(6)}_{1,1}\right]
    \right]
    +
    {\cal O}(a^2).
\end{align}
The corresponding frequency equation is
\begin{equation}
    \omega_{\rm P,Taylor}^{(6)}(a)^2
    =
    {\cal V}_0(\omega_{\rm P,Taylor}^{(6)},a)
    -
    \left[
    2\hhbar\,
    {\cal P}^{(6)}_{\rm P}
    (\omega_{\rm P,Taylor}^{(6)},a;\hhbar)
    \right]_{\hhbar=i}.
\end{equation}
The first label, ${\rm P}$, denotes Pad\'e resummation of the WKB series, while the second label, ${\rm Taylor}$, denotes that the spin dependence has not yet been Pad\'e resummed. This equation is implicit because the right-hand side depends on $\omega_{\rm P,Taylor}^{(6)}$.

To see how the coefficient matching works, solve this equation with
\begin{equation}
    \omega_{\rm P,Taylor}^{(6)}(a)
    =
    \omega_0+\omega_1a+{\cal O}(a^2).
\end{equation}
Substituting this ansatz into the implicit equation above, moving all terms to one side, and expanding to first order in $a$ gives
\begin{align}
    0
    &=
    \omega_0^2
    -
    \left(0.131718721376
    -
    0.066484876685\,{\rm i}\right)
    \nonumber \\
    &
    +
    a
    \bigg[
    2\omega_0\omega_1
    -
    \left(0.047175943281
    -
    0.010443242038\,{\rm i}\right)
    \bigg]
    \nonumber \\
    &
    +
    {\cal O}(a^2).
\end{align}
Thus,
\begin{align}
    \omega_0
    &=
    0.373674677244
    -
    0.088960907353\,{\rm i},
    \\
    \omega_1
    &=
    0.062886808557
    +
    0.000997783777\,{\rm i}.
\end{align}

Repeating this coefficient-matching procedure through sixth order in the spin gives the WKB-Pad\'e-resummed slow-rotation series
\begin{align}
    \omega_{\rm P,Taylor}^{(6)}(a)
    &=
    \left(
    0.373674677244
    -
    0.088960907353\,{\rm i}
    \right)
    \nonumber \\
    &
    +
    \left(
    0.062886808557
    +
    0.000997783777\,{\rm i}
    \right)a
    \nonumber \\
    &
    +
    \left(
    0.044794283654
    +
    0.006200457803\,{\rm i}
    \right)a^2
    \nonumber \\
    &
    +
    \left(
    0.020972766467
    +
    0.002566343431\,{\rm i}
    \right)a^3
    \nonumber \\
    &
    +
    \left(
    0.016907115331
    +
    0.004518948094\,{\rm i}
    \right)a^4
    \nonumber \\
    &
    +
    \left(
    0.010519133990
    +
    0.002461836977\,{\rm i}
    \right)a^5
    \nonumber \\
    &
    +
    \left(
    0.008339909678
    +
    0.002329847106\,{\rm i}
    \right)a^6 \nonumber \\
   & +
    {\cal O}(a^7).
\end{align}

The expression above is still a Taylor series in the spin. We now perform the second Pad\'e resummation, this time in $a$. The diagonal spin-Pad\'e approximant is
\begin{equation}
    \omega_{\rm P,P}^{(6)}(a)
    =
    \frac{
    u_0+u_1a+u_2a^2+u_3a^3
    }{
    1+v_1a+v_2a^2+v_3a^3
    }.
\end{equation}
The notation ${\rm P,P}$ means that the WKB series has first been Pad\'e resummed in $\hhbar$, and that the resulting spin series has then been Pad\'e resummed in $a$. The coefficients are fixed by requiring the Taylor expansion of $\omega_{\rm P,P}^{(6)}(a)$ about $a=0$ to agree with $\omega_{\rm P,Taylor}^{(6)}(a)$ through ${\cal O}(a^6)$. This gives
\begin{align}
    u_0
    &=
    0.373674677244
    -
    0.088960907353\,{\rm i},
    \\
    u_1
    &=
    -
    0.141440143288
    +
    0.123805471847\,{\rm i},
    \\
    u_2
    &=
    -
    0.104944879466
    +
    0.017860120533\,{\rm i},
    \\
    u_3
    &=
    0.021434402261
    -
    0.022678605347\,{\rm i},
\end{align}
and
\begin{align}
    v_1
    &=
    -
    0.591520024402
    +
    0.187825224066\,{\rm i},
    \\
    v_2
    &=
    -
    0.284807443539
    -
    0.066631717931\,{\rm i},
    \\
    v_3
    &=
    0.131800184480
    -
    0.036907113848\,{\rm i}.
\end{align}
Therefore, the double-resummed illustrative result is
\begin{align}
    \omega_{\rm P,P}^{(6)}(a)
    &=
    \frac{
    u_0+u_1a+u_2a^2+u_3a^3
    }{
    1+v_1a+v_2a^2+v_3a^3
    }.
\end{align}
By construction, expanding this rational function in $a$ reproduces the single-resummed slow-rotation series $\omega_{\rm P,Taylor}^{(6)}(a)$ through ${\cal O}(a^6)$. The calculation in the main text follows the same steps, but uses higher orders in both the $a$ series and the WKB series, described in Sec.~\ref{sec: slow-rotation}. If the WKB series is Borel-Pad\'e resummed instead of Pad\'e resummed, the analogous quantities are denoted by $\omega_{\rm BP,Taylor}$ before the spin resummation and by $\omega_{\rm BP,P}$ after the spin resummation.

\section{Near-extremal near-horizon Chandrasekhar-Detweiler potential} 
\label{app: CD potential}

This appendix gives the full algebraic expression for the term omitted from Eq.~(\ref{eq: NHEK CD potential}) in the near-extremal, near-horizon expansion of the Chandrasekhar-Detweiler potential.

The third term at $\mathcal{O}(1)$ in Eq.(\ref{eq: NHEK CD potential}) is
\bw
\begin{equation}
-\frac{\big(r-r_-\big)\big(r-r_+\big)\,A\,B}{32\big(r-r_{\omega+}\big)^4\big[3(r-1)+(r-r_{\omega+})\lambda\big]^2\big\{-3(r-r_-)(r-r_+)+2(r-r_{\omega+})\big[3(r-1)+(r-r_{\omega+})\lambda\big]\big\}},
\end{equation}
where
\begin{equation}
\begin{split}
A =& 6(r-1)\Big\{(r-r_-)\big[(r-r_+)-(r-r_{\omega+})\big]-(r-r_+)(r-r_{\omega+})\Big\}\\
&-(r-r_{\omega+})\Big[(r-r_-)(r-r_+)(-6+\kappa_2-4\lambda)+2(r-r_-)(r-r_{\omega+})\lambda+2(r-r_+)(r-r_{\omega+})\lambda\Big],
\end{split}
\end{equation}
\begin{equation}
\begin{split}
B =& 6(r-1)\Big\{3(r-r_-)\big[(r-r_+)-(r-r_{\omega+})\big]+(r-r_{\omega+})\big(-3(r-r_+)+2(r-r_{\omega+})\kappa_2\big)\Big\}\\
&+(r-r_{\omega+})\Big\{2(r-r_{\omega+})\big[-3(r-r_+)+2(r-r_{\omega+})\kappa_2\big]\lambda-3(r-r_-)\big[(r-r_+)(-6+\kappa_2-4\lambda)+2(r-r_{\omega+})\lambda\big]\Big\},
\end{split}
\end{equation}
and $\kappa_2$ and $\lambda$ are constants evaluated at the extremal limit $a=1$.
\ew

\section{Eikonal limit of the Chandrasekhar-Detweiler potential} 
\label{app: eikonal}

This appendix records the eikonal limit of the Chandrasekhar-Detweiler potential and shows how the leading-order WKB condition recovers the standard eikonal relation for Kerr QNM frequencies.

The Chandrasekhar-Detweiler potential in the eikonal limit ($l\gg1$) is
\begin{equation}
    V=\frac{(\lambda+2)\Delta}{\bar{\rho}^4}.
\end{equation}
The condition for the peak $\frac{d V}{d r}|_{r=r_0}=0$ gives
\begin{equation}
   (r_0-1) a m + (r_0-3) r_0^2 \omega + (r_0+1)a^2\omega=0.
\end{equation}
Assuming $\Re[\omega]\gg\Im[\omega]$ in the eikonal limit, we have
\begin{equation}
    \Re[\omega]=\left(l+\frac{1}{2}\right)\frac{(r_0-1)\mu a}{ (3-r_0) r_0^2 - (r_0+1)a^2},
\end{equation}
where $\mu = m/(l+\frac{1}{2})$. 
The first-order WKB formula gives
\begin{equation}
    \Re[\omega] = \sqrt{V}|_{\omega=\Re[\omega],r=r_0}.
\end{equation}
Taking the eikonal limit, we find
\begin{equation}
\begin{split}
2r_0^4(r_0 - 3)^2 + 4r_0^2\left[(1-\mu^2)r_0^2 - 2r_0 - 3(1-\mu^2)\right]a^2 \\
+ (1-\mu^2)\left[(2-\mu^2)r_0^2 + 2(2+\mu^2)r_0 + (2-\mu^2)\right]a^4=0.
\end{split}
\end{equation}
These agree with the leading-order eikonal WKB conditions of Ref.~\cite{Yang:2012he}, showing that the Chandrasekhar-Detweiler potential reproduces the expected eikonal conditions in this limit.
However, when the eikonal limit $l\gg1$ is not applied, the radial Teukolsky equation becomes a complex function of $r$, where the physical interpretations of the peak are less clear.

\bibliographystyle{apsrev4-1}
\bibliography{Ref}
\end{document}